\RequirePackage[mathlines]{lineno} 
\documentclass[a4paper,11pt]{article}

\usepackage{jheppub} 
\usepackage{overpic}
\usepackage{amssymb}
\usepackage{amsmath}
\usepackage[T1]{fontenc} 
\usepackage{amsmath}
\usepackage{graphicx}
\usepackage{subfigure}
\usepackage{epstopdf}
\usepackage{rotating}

\usepackage{threeparttable}
\usepackage{multirow}
\usepackage{subfigure}
\usepackage{float}
\usepackage{verbatim}

\let\oldequation\equation
\let\oldendequation\endequation
\renewenvironment{equation}
  {\linenomathNonumbers\oldequation}
  {\oldendequation\endlinenomath}

  \newcommand{\etalnu}{D^+ \to \eta\ell^+\nu_{\ell}}
  \newcommand{\etaenu}{D^+\to \eta e^+\nu_e}
  \newcommand{\etamunu}{D^+\to \eta\mu^+\nu_\mu}

  \newcommand{\ffeta}{f^\eta_+(0)}

\begin{document}


\title{\bf \boldmath \texorpdfstring{
Improved Measurements of $D^+ \to \eta e^+\nu_e$ and $D^+ \to \eta \mu^+\nu_\mu$}{}
}

\collaborationImg{\includegraphics[height=4cm,angle=0 ]{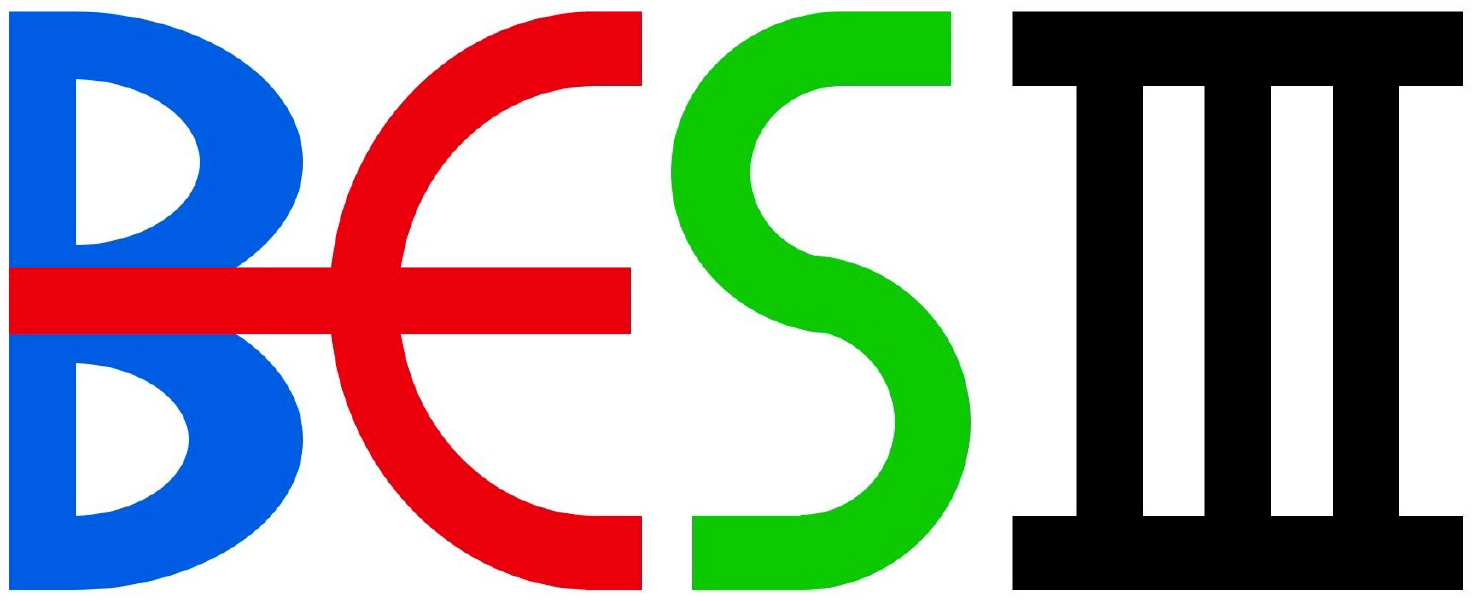}}
\collaboration{BESIII Collaboration}

 \newcommand{\BESIIIorcid}[1]{\href{https://orcid.org/#1}{\hspace*{0.1em}\raisebox{-0.45ex}{\includegraphics[width=1em]{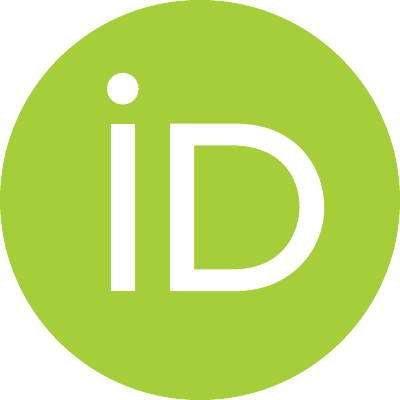}}}}
\author{
M.~Ablikim$^{1}$\BESIIIorcid{0000-0002-3935-619X},
M.~N.~Achasov$^{4,c}$\BESIIIorcid{0000-0002-9400-8622},
P.~Adlarson$^{77}$\BESIIIorcid{0000-0001-6280-3851},
X.~C.~Ai$^{82}$\BESIIIorcid{0000-0003-3856-2415},
R.~Aliberti$^{36}$\BESIIIorcid{0000-0003-3500-4012},
A.~Amoroso$^{76A,76C}$\BESIIIorcid{0000-0002-3095-8610},
Q.~An$^{59,73,a}$,
Y.~Bai$^{58}$\BESIIIorcid{0000-0001-6593-5665},
O.~Bakina$^{37}$\BESIIIorcid{0009-0005-0719-7461},
Y.~Ban$^{47,h}$\BESIIIorcid{0000-0002-1912-0374}
% ...
\\
\vspace{0.2cm}
(BESIII Collaboration)\\
\vspace{0.2cm} {\it
$^{1}$ Institute of High Energy Physics, Beijing 100049, People's Republic of China\\
% ...
$^{4}$ Budker Institute of Nuclear Physics SB RAS (BINP), Novosibirsk 630090, Russia\\
% ...
$^{36}$ Johannes Gutenberg University of Mainz, Johann-Joachim-Becher-Weg 45, D-55099 Mainz, Germany\\
$^{37}$ Joint Institute for Nuclear Research, 141980 Dubna, Moscow region, Russia\\
% ...
$^{47}$ Peking University, Beijing 100871, People's Republic of China\\
% ...
$^{58}$ Southeast University, Nanjing 211100, People's Republic of China\\
$^{59}$ State Key Laboratory of Particle Detection and Electronics, Beijing 100049, Hefei 230026, People's Republic of China\\
% ...
$^{73}$ University of Science and Technology of China, Hefei 230026, People's Republic of China\\
% ...
$^{76}$ University of Turin and INFN, (A)University of Turin, I-10125, Turin, Italy; (B)University of Eastern Piedmont, I-15121, Alessandria, Italy; (C)INFN, I-10125, Turin, Italy\\
$^{77}$ Uppsala University, Box 516, SE-75120 Uppsala, Sweden\\
% ...
$^{82}$ Zhengzhou University, Zhengzhou 450001, People's Republic of China\\

\vspace{0.2cm}
$^{a}$ Deceased\\
% ...
$^{c}$ Also at the Novosibirsk State University, Novosibirsk, 630090, Russia\\
% ...
$^{h}$ Also at State Key Laboratory of Nuclear Physics and Technology, Peking University, Beijing 100871, People's Republic of China\\
% ...

}
}

\abstract{
Using 20.3 fb$^{-1}$ of $e^+e^-$ collision data collected at the center-of-mass energy of 3.773 GeV with the BESIII detector, we measure the branching fractions of $\etaenu$ and $\etamunu$ to be $(9.75\pm0.29\pm0.28)\times10^{-4}$ and $(9.08\pm0.35\pm0.23)\times10^{-4}$, where the first and second uncertainties are statistical and systematic, respectively. From a simultaneous fit to their partial decay rates, we determine the product of the hadronic form factor $\ffeta$ and the modulus of the $c\to d$ Cabibbo-Kobayashi-Maskawa matrix element $|V_{cd}|$ to be $\ffeta|V_{cd}|=0.078\pm0.002\pm0.001$. Taking the $|V_{cd}|$ value from the Standard Model global fit as input, we obtain $\ffeta=0.345\pm0.008\pm0.003$. The ratio between the measured branching fractions of $D^+\to\eta^+\mu^+\nu_{\mu}$ and $D^+\to\eta e^+\nu_e$, is determined to be $0.93\pm0.05_{\rm stat.}\pm0.02_{\rm syst.}$, indicating no violation of lepton flavor universality.
}

\maketitle
\flushbottom

\section{INTRODUCTION}

Experimental studies of the semileptonic decays of charmed mesons are crucial for exploring the weak and strong interactions in the charm sector~\cite{Ke:2023qzc}.
By analyzing their decay dynamics, one can extract the product of the modulus of the Cabibbo-Kobayashi-Maskawa (CKM) matrix element $|V_{cd(s)}|$ and the hadronic transition form factor.
For the $D^+\to \eta\ell^+\nu_\ell$ decay, where $\ell$ denotes either an electron or a muon, the hadronic transition form factor at zero-momentum transfer $f^\eta_+(0)$ can be calculated via several theoretical approaches, e.g., QCD light-cone sum rules (LCSR)~\cite{dse:2015gdu,dse:2013nof}, the covariant light-front quark model (LFQM)~\cite{dse:2012rcv}, and the covariant confined quark model (CCQM) ~\cite{dse:2018nrs,dse:2018nrsnew}. The predicted values of $f^\eta_+(0)$ are summarized in Table~\ref{tab:FF}.
Using the value of $|V_{cd}|$ provided by the CKM Fitter group~\cite{pdg2022}, the hadronic transition form factor can be extracted, resulting in a stringent test of the theoretical predictions.
Alternatively, using the predicted value of $f^\eta_+(0)$ from theory as input allows the extraction of $|V_{cd}|$, which is important for testing the unitarity of the CKM matrix. Scrutinizing the branching fraction (BF) ratios of SL $D^+$ decays could also probe the lepton flavor universality (LFU), and therefore test the standard model (SM) in the charm sector.

\begin{table}[htp]
	\centering
	\caption{\label{tab:FF}
		\small   Theoretical predictions of the hadronic transition form factor of $f^\eta_+(0)$.}
	\begin{tabular}{lc} \hline\hline
		Approach &$f_+^{\eta}(0)$\\ \hline
		LCSR~\cite{dse:2015gdu}                      &$0.429^{+0.165}_{-0.141}$\\
		LCSR~\cite{dse:2013nof}                      &$0.552\pm0.051$\\
		LFQM~\cite{dse:2012rcv}                      & $0.71^{-0.00+0.01}_{+0.00-0.01}$       \\
		CCQM~\cite{dse:2018nrs}                      &$0.67\pm0.11$\\
		CCQM~\cite{dse:2018nrsnew}                   &$0.36\pm0.05$\\
		\hline\hline
	\end{tabular}
\end{table}

The form factor $f^\eta_+(0)$ and the BFs of $D^+ \to \eta e^+\nu_e$ and $D^+ \to \eta\mu^+\nu_\mu$ have been measured by CLEO~\cite{cleo:2011jy} and BESIII~\cite{bes3:zhangyu,bes3:Detamuv}.
The previous BESIII measurements were based on 2.93 fb$^{-1}$ of $e^+e^-$ collision data taken at the center-of-mass energy $\sqrt{s} = 3.773~\text{GeV}$ during 2010 and 2011.
This paper reports improved measurements of the BFs, decay dynamics and LFU test of $D^+ \to \eta e^+\nu_e$ and $D^+ \to \eta\mu^+\nu_\mu$ using 20.3~fb$^{-1}$~\cite{20fb} of $e^+e^-$ collision data collected by the BESIII detector at $\sqrt s=3.773$ GeV in 2010-2011 and 2022-2024. The $\eta$ mesons are reconstructed through the decay modes of $\eta\to\gamma\gamma$ and $\eta\to\pi^+\pi^-\pi^0$. The dataset of 20.3~fb$^{-1}$ includes the previous analyzed 2.93~fb$^{-1}$ subset.
Throughout this paper, charge-conjugated modes are implied.

\section{BESIII DETECTOR AND MONTE CARLO SIMULATION}

The BESIII detector~\cite{Ablikim:2009aa} records symmetric $e^+e^-$ collisions
provided by the BEPCII storage ring~\cite{Yu:IPAC2016-TUYA01} in the center-of-mass energy range from 1.85 to 4.95~GeV, with a peak luminosity of $1.1 \times 10^{33}\;\text{cm}^{-2}\text{s}^{-1}$
achieved at $\sqrt{s} = 3.773\;\text{GeV}$.
BESIII has collected large data samples in this energy region~\cite{Ablikim:2019hff,Li:2021iwf}. The cylindrical core of the BESIII detector covers 93\% of the full solid angle and consists of a helium-based
multilayer drift chamber~(MDC), a plastic scintillator time-of-flight
system~(TOF), and a CsI(Tl) electromagnetic calorimeter~(EMC),
which are all enclosed in a superconducting solenoidal magnet
providing a 1.0~T magnetic field. The solenoid is supported by an
octagonal flux-return yoke with resistive plate counter muon
identification modules embedded in steel.
The charged-particle momentum resolution at $1~{\rm GeV}/c$ is
$0.5\%$, and the ${\rm d}E/{\rm d}x$ resolution is $6\%$ for electrons
from Bhabha scattering. The EMC measures photon energies with a
resolution of $2.5\%$ ($5\%$) at $1$~GeV in the barrel (end cap)
region. The time resolution in the TOF barrel region is 68~ps, while
that in the end cap region was 110~ps. The end cap TOF
system was upgraded in 2015 using multi-gap resistive plate chamber
technology, improving the time resolution to 60~ps~\cite{etof}.
Approximately 86\% of the data used in this analysis were collected after this upgrade.

Monte Carlo (MC) simulated data samples produced with a {\sc
	geant4}-based~\cite{geant4} software package, which
includes the geometric description of the BESIII detector and the
detector response, are used to determine detection efficiencies
and to estimate the background contributions. The simulation models the beam
energy spread and initial state radiation (ISR) in the $e^+e^-$
annihilations with the {\sc kkmc} generator~\cite{ref:kkmc}.
The inclusive MC sample includes the production of $D\bar{D}$
pairs (including quantum coherence for the neutral $D$ channels),
	the non-$D\bar{D}$ decays of the $\psi(3770)$, the ISR
	production of the $J/\psi$ and $\psi(3686)$ states, and the
	continuum processes incorporated in {\sc kkmc}~\cite{ref:kkmc}.
All particle decays are modeled with {\sc
	evtgen}~\cite{ref:evtgen} using BFs
either taken from the
Particle Data Group~\cite{pdg2022}, when available,
otherwise estimated with {\sc lundcharm}~\cite{ref:lundcharm}.

\section{DATA ANALYSIS METHOD}

At $\sqrt s=3.773$~GeV, the $D$ and $\bar D$ mesons are produced in pair from the $e^+e^-\to \psi(3770)\to D\bar D$ process, where $D$ denotes $D^0$ or $D^+$.
This feature enables absolute BF measurements using the well-established double-tag (DT) method.
In this method, the single-tag (ST) candidate events are selected by reconstructing a $D^-$ with the hadronic final states
$D^- \to K^{+}\pi^{-}\pi^{-}$, $K^0_{S}\pi^{-}$, $K^{+}\pi^{-}\pi^{-}\pi^{0}$, $K^0_{S}\pi^{-}\pi^{0}$, $K^0_{S}\pi^{+}\pi^{-}\pi^{-}$, and $K^{+}K^{-}\pi^{-}$.
These selected candidates are referred to as ST $D^-$ candidates.
In the presence of the ST $D^-$ mesons, signal candidates
are selected from the remaining particles to form the DT events.
The BF of the signal decay is determined by
\begin{equation}
	\label{eq:bf}
	{\mathcal B}_{\rm sig}=N_{\rm DT}/(N_{\rm ST}^{\rm tot}\cdot \bar{\epsilon}_{\rm sig}),
\end{equation}
where $N_{\rm ST}^{\rm tot}$ and $N_{\rm DT}$ are the total yield of ST $D^-$ mesons and the yield of DT events in data,
$\bar{\epsilon}_{\rm sig}$ is the average signal efficiency weighted by the ST yields of the $i$-th tag mode in data,
\begin{equation}
	\bar{\epsilon}_{\rm sig} = \sum_i(N_{\rm ST}^{i}\cdot\epsilon^{i}_{\rm sig})/N^{\rm tot}_{\rm ST}=\sum_i(N_{\rm ST}^{i}\cdot\epsilon^{i}_{\rm DT}/\epsilon^{i}_{\rm ST})/N^{\rm tot}_{\rm ST},
\end{equation}
where $N_{\rm ST}^{i}$ is the number of ST $D^-$ mesons for the $i$-th tag mode in data, $\epsilon^i_{\rm sig}$ is the signal efficiency of the $i$-th tag mode,  $\epsilon^i_{\rm ST}$ is the efficiency of reconstructing the ST mode $i$, $\epsilon^i_{\rm DT}$ is the efficiency of finding the tag mode $i$ and the semileptonic decay simultaneously.

\section{SINGLE-TAG $D^-$ CANDIDATES}

\begin{itemize}

	\item {\bf Good charged track}\\
	\begin{itemize}
		\item Charged tracks detected in the MDC are required to be within a polar angle ($\theta$) range of $|\rm{cos\theta}|<0.93$, where $\theta$ is defined with respect to the $z$-axis,
		which is the symmetry axis of the MDC.

		\item For charged tracks not originating from $K_S^0$ decay, the distance of closest approach to the interaction point (IP)
		must be less than 10\,cm
		along the $z$-axis, $|V_{z}|$,
		and less than 1\,cm
		in the transverse plane, $|V_{xy}|$.
	\end{itemize}

	\item {\bf Good photon selection}\\
	\begin{itemize}
		\item Photon candidates are identified using isolated showers in the EMC.  The deposited energy of each shower must be greater than 25~MeV in the barrel region ($|\cos \theta|< 0.80$) and greater than 50~MeV in the end cap region ($0.86 <|\cos \theta|< 0.92$).

		\item
		To exclude showers that originate from charged tracks, the opening angle between the shower direction and the extrapolated position on the EMC of the nearest charged track is required to be no less than 10 degrees.

\item  To suppress electronic noise and showers unrelated to the event, the difference between the EMC time and the event start time is required to be within
[0, 700]\,ns.
\end{itemize}

\item {\bf Particle identification for hadrons}
\begin{itemize}
\item  Particle identification~(PID) for charged tracks combines measurements of the energy deposited in the MDC~(d$E$/d$x$) and the flight time in the TOF to form likelihoods $\mathcal{L}(h)~(h=K,\pi)$ for each hadron $h$ hypothesis.
Tracks are identified as kaons and pions by comparing the likelihoods for the kaon and pion hypotheses, $\mathcal{L}(K)>\mathcal{L}(\pi)$ and $\mathcal{L}(\pi)>\mathcal{L}(K)$, respectively.
\end{itemize}

\item {\bf $\pi^{0}$ reconstruction}

The $\pi^{0}$ mesons are reconstructed through the
$\pi^{0}\to\gamma\gamma$ decay. The $\gamma\gamma$ combinations
satisfying the following selection criteria are regarded
as a $\pi^{0}$ meson:
\begin{itemize}
	\item
	Two good photons must pass the good photon selection criteria as described above;
	\item
	$\pi^0$ mass window: 0.115 GeV/$c^{2}<M_{\gamma\gamma}<0.150$ GeV/$c^{2}$,
	where $M_{\gamma\gamma}$ is the invariant mass of the $\gamma\gamma$ combinations;
	\item
	The $\gamma\gamma$ invariant mass is constrained to
	the $\pi^{0}$ nominal mass~\cite{pdg2022} by a one-constraint (1-C) kinematic fit to improve its momentum resolution and
	the momenta from the kinematic fit are kept for further analysis.  For any of the $\pi^0$ candidates, the $\chi^2$ of the 1-C kinematic fit is required to be less than 50.
\end{itemize}

\item {\bf Secondary vertex for $K_{S}^0$}
\begin{itemize}
\item Each $K_{S}^0$ candidate is reconstructed from two oppositely charged tracks satisfying $|V_{z}|<$ 20~cm.
The two charged tracks are assigned
as $\pi^+\pi^-$ without imposing further PID criteria. They are constrained to
originate from a common vertex and are required to have an invariant mass
within $|M_{\pi^{+}\pi^{-}} - m_{K_{S}^{0}}|<$ 12~MeV$/c^{2}$, where
$m_{K_{S}^{0}}$ is the $K^0_{S}$ nominal mass~\cite{pdg2022}. The
decay length of the $K^0_S$ candidate is required to be greater than
twice the vertex resolution away from the IP.
\end{itemize}
\end{itemize}

To identify the ST $D^-$ mesons, we use two kinematic variables: the energy difference $\Delta E\equiv E_{D^-}-E_{\rm beam}$ and the beam-constrained mass $M_{\rm BC}\equiv\sqrt{E_{\rm beam}^2-|\vec{p}_{D^-}|^2}$, where $E_{\rm beam}$ is the beam energy, and $E_{D^-}$ and $\vec{p}_{D^-}$ are the total energy and momentum of the ST $D^-$ meson in the $e^+e^-$ center-of-mass frame.
If multiple $D^-$ candidates exist for a specific tag mode, the one giving the smallest $|\Delta E|$ is chosen for the further analyses.
To suppress combinatorial background in the $M_{\rm BC}$ distribution, tag-dependent $\Delta E$ requirements are imposed on the ST candidates.
The detailed $\Delta E$ requirements and the efficiencies of detecting ST $D^-$ mesons estimated by analyzing the inclusive MC sample are summarized in Table~\ref{ST:realdata}.

\begin{figure}[htbp]\centering
		\includegraphics[width=1.0\linewidth]{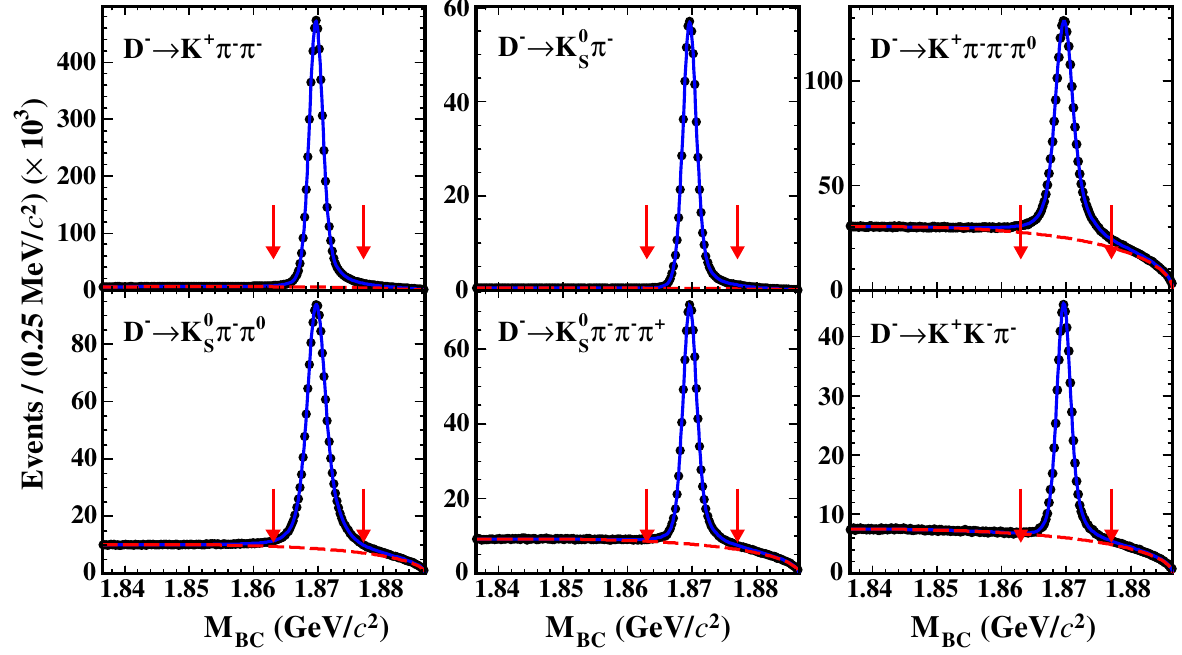}
		\caption{
			Fits to the $M_{\rm BC}$ distributions of the ST $D^-$ candidates.
			The points with error bars are data, the blue curves are the best fits, and the red dashed curves are the fitted combinatorial background shapes. The pairs of red arrows show the $M_{\rm BC}$ signal window.
			\label{fig:datafit_Massbc}}
	\end{figure}

For each tag mode, the yield of ST $D^-$ mesons is extracted by an unbinned maximum likelihood fit to the corresponding $M_{\rm BC}$ distribution.
In the fit, the MC-simulated signal shape is convolved with a double-Gaussian function to account for the resolution difference between data and MC simulation, and all parameters are float; while the background shape is parameterized as an ARGUS function~\cite{argus} with its endpoint fixed at the $E_{\rm beam}=1.8865~{\rm GeV/c^2}$. 
Figure~\ref{fig:datafit_Massbc} shows the results of the fits to the $M_{\rm BC}$ distributions of the accepted ST candidates in data for individual tag modes. The candidates with $M_{\rm BC}\in (1.863,1.877)$ GeV/$c^2$ for $D^-$ tags are kept for further analyses. Summing over the tag modes yields a total $(10677.9\pm3.8_{\rm stat})\times 10^3$ ST $D^-$ mesons.

\begin{table}
	\renewcommand{\arraystretch}{1.2}
	\centering
	\caption {The $\Delta E$ requirements, the obtained ST $D^-$ yields in data ($N_{\rm ST}^{i}$), and the efficiencies of detecting ST $D^-$ mesons ($\epsilon_{\rm ST}^{i}$). The uncertainties are statistical only.           \label{ST:realdata}}
			\begin{tabular}{c|c|c|c}
				\hline
				\hline
				$D^-$ mode & $\Delta E$~(MeV)  &  $N_{\rm ST}^{i}~(\times 10^3)$  &  $\epsilon_{\rm ST}^{i}~(\%)    $       \\\hline
				$K^+\pi^-\pi^-$                  &  $(-25,24)$ & $5567.2\pm2.5$&$51.08\pm0.00$\\
				$K^{0}_{S}\pi^{-}$               &  $(-25,26)$ & $656.5\pm0.8$&$51.42\pm0.01$\\
				$K^{+}\pi^{-}\pi^{-}\pi^{0}$     &  $(-57,46)$ & $1740.2\pm1.9$&$24.53\pm0.00$\\
				$K^{0}_{S}\pi^{-}\pi^{0}$        &  $(-62,49)$ & $1442.4\pm1.5$&$26.45\pm0.00$\\
				$K^{0}_{S}\pi^{-}\pi^{-}\pi^{+}$ &  $(-28,27)$ & $790.2\pm1.1$&$29.68\pm0.01$\\
				$K^{+}K^{-}\pi^{-}$              &  $(-24,23)$ & $481.4\pm0.9$&$40.91\pm0.01$\\
				\hline
				\hline
			\end{tabular}
	\end{table}

\section{DOUBLE-TAG EVENTS}

The candidates for $\etaenu$, and $\etamunu$ are selected from the remaining tracks in the signal side, given a tagged $D^-$ candidate.
Candidates for charged tracks, photons, and $\pi^0$ are selected with the same criteria as in the ST selection. To select $\eta$ candidates from both decay modes, the following criteria are applied: 
For the $\eta \to \gamma\gamma$ decay, the two-photon invariant mass must fall within (0.50, 0.57)~GeV/$c^2$, while for the $\eta \to \pi^+\pi^-\pi^0$ decay, the invariant mass is required to be in the range (0.53, 0.57)~GeV/$c^2$~\cite{bes3:zhangyu}. 
A 1-C kinematic fit is performed for both decay channels to constrain the reconstructed mass to the nominal $\eta$ mass value, with a $\chi^2$ threshold of 50 required for the $\gamma\gamma$ mode. 
In events with multiple candidates per decay channel, only the candidate with the smallest $\chi^2$ value is retained for further analysis. The positron and muon are identified using the TOF, $dE/dx$, and EMC measurements with the combined confidence levels $\mathcal{L}{_e}$, $\mathcal{L}{_\mu}$, $\mathcal{L}{_K}$, and $\mathcal{L}{_\pi}$, which are calculated for electron, muon, kaon and pion hypotheses, respectively.
The positron candidate is required to satisfy $\mathcal{L}{_e} > 0.8\times (\mathcal{L}{_e}+\mathcal{L}{_\pi}+\mathcal{L}{_K})$, and $\mathcal{L}{_e} >0.001$. Furthermore, the energy of the positron deposited in the EMC divided by the track momentum is required
to be larger than 0.8 for $D^+\to\eta e^+\nu_e$.
The muon candidate is required to satisfy $\mathcal{L}{_\mu} > \mathcal{L}{_K}$, $\mathcal{L}{_\mu} > \mathcal{L}{_e}$, and $\mathcal{L}{_\mu} >0.001$, and the energy of muon deposited in the EMC
is required to be within $(0.101,0.282)$ GeV.  We determine this requirement by maximizing the value of $S/\sqrt{S+B}$, where $S$ and $B$ denote the signal and background yields based on the inclusive MC sample.

To suppress the backgrounds from hadronic $D^+$ decays, no additional good charged tracks ($N_{\rm extra}^{\rm char}$) are allowed in the signal side.
To reject the backgrounds from the hadronic decays involving $\pi^0$, the maximum energies of extra photons ($E_{\text{extra~}\gamma}^{\rm max}$) which have not been used in the event selection are required to be less than 0.25~GeV for all signal modes, and no extra $\pi^0$ mesons ($N_{\text{extra~}}^{\pi^0}$) for $D^+\to\eta\mu^+\nu_\mu$ is allowed. To suppress the peaking background of $D^+\to\eta\pi^+$, the invariant mass of $\eta\mu^+$ ($M_{\eta\mu^+}$) is required to be less than 1.722~${\rm GeV}/c^2$ for $D^+\to\eta_{(2\gamma)}\mu^+\nu_\mu$ and less than 1.710~${\rm GeV}/c^2$ for $D^+\to\eta_{(3\pi)}\mu^+\nu_\mu$.

Since neutrino is not detectable and its mass is nearly zero, the kinematics of semileptonic $D^+$ candidates are characterized by the variable of $U_{\rm miss}\equiv E_{\rm miss}-|\vec{p}_{\rm miss}|c$, which is expected to peak around 0. The $E_{\rm miss}$ and $\vec{p}_{\rm miss}$, given by $E_{\rm miss}\equiv E_{\rm beam}-E_{\eta}-E_{\ell^+}$ ($\ell=e, \mu$) and $\vec{p}_{\rm miss}\equiv\vec{p}_{D^+}-\vec{p}_{\eta}-\vec{p}_{\ell^+}$, are the missing energy and momentum of the DT event in the $e^+e^-$ center-of-mass frame, respectively. Here, the $E_{\eta(\ell^+)}$ and $\vec{p}_{\eta(\ell^+)}$ are the measured energy and momentum of the $\eta(\ell^+)$ candidates, respectively, and $\vec{p}_{D^+}\equiv-\hat{p}_{D^-} \sqrt{E_{\rm beam}^2/c^2-m_{D^-}^2 c^2 }$, where $\hat{p}_{D^-}$ is the unit vector in the momentum direction of the ST $D^-$ meson and $m_{D^-}$ is the nominal $D^-$ mass~\cite{pdg2022}.
The beam energy and the nominal $D^-$ mass are used in the calculation of the ST $D^-$ momentum, thereby improving the $U_{\rm miss}$ resolution.

\section{DETERMINATION OF BRANCHING FRACTIONS}

\subsection{Signal Efficiencies and Branching Fractions}

\begin{figure}[htbp]
	\begin{center}\centering
		\includegraphics[width=0.72\linewidth]{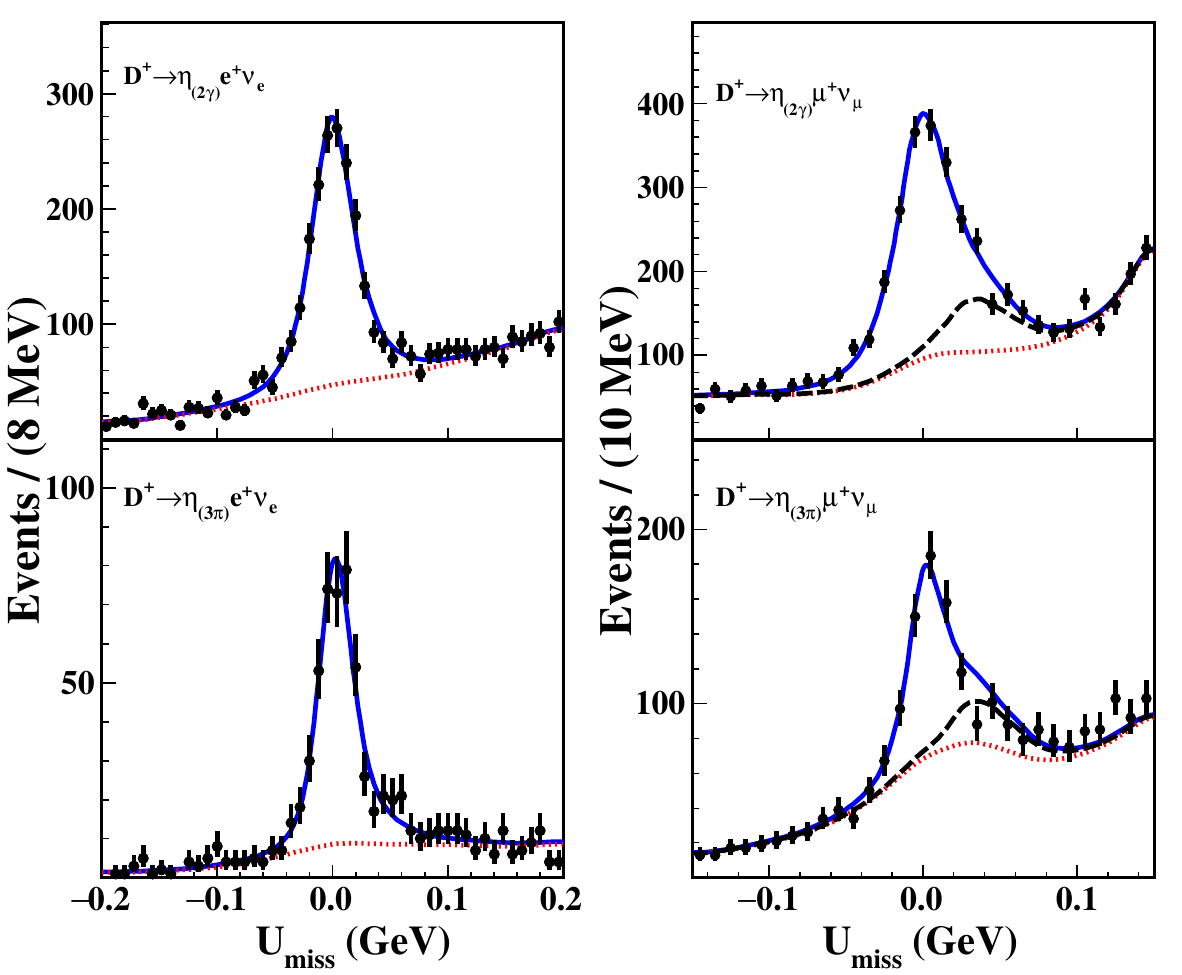}
		\caption{
			Fits to the $U_{\rm miss}$ distributions of the candidates for different signal decays.
			The dots with error bars are data,
			The blue solid lines denote the total fit.
			The black dashed lines denote the fitted peaking backgrounds, and
			the red dashed lines are the fitted combinatorial background shapes.
			\label{fit_umiss}
		}
	\end{center}
\end{figure}

After imposing all selection criteria, the $U_{\rm miss}$ distributions of the accepted candidates for the semileptonic decays are obtained, as shown in Fig.~\ref{fit_umiss}. For each signal decay, the signal yield is derived from an unbinned maximum likelihood fit to the corresponding $U_{\rm miss}$ distribution.
In the fit, the signals are modeled by the MC-simulated shape extracted from the inclusive MC sample, convolved with a Gaussian function to account for the resolution difference between data and MC simulation, and all parameters are float. The combinatorial backgrounds are described by the shapes derived from inclusive MC simulation, and their yields are left free.
In the fits to $D^+\to\eta\mu^+\nu_\mu$, the shapes of peaking backgrounds from $D^+\to\eta\pi^+\pi^0$ are derived from inclusive MC, and the yields are fixed to be 389 and 151 according to the misidentification rate and the BFs in the PDG for $D^+\to\eta_{(2\gamma)}\mu^+\nu_\mu$ and $D^+\to\eta_{(3\pi)}\mu^+\nu_\mu$, respectively.

\begin{table}[htbp]
	\centering\linespread{1.1}
	\caption{
		The obtained DT efficiencies ($\epsilon_{\rm DT,\eta\ell^+\nu_\ell}$) and signal efficiencies ($\epsilon_{\eta\ell^+\nu_\ell}$) in \% for different signal decays in each tag mode. The listed efficiencies do not include the BFs of the sub-resonant decays. The uncertainties are statistical. }
	\small
	\label{tab:bf}
	\resizebox{1.0\textwidth}{!}{
		\begin{tabular}{c c c  cc cccc }\hline \hline
			Tag mode &$\epsilon_{{\rm DT}, \eta_{(2\gamma)}e^+\nu_e}$&$\epsilon_{\eta_{(2\gamma)}e^+\nu_e}$&$\epsilon_{{\rm DT},\eta_{(3\pi)}e^+\nu_e}$&$\epsilon_{\eta_{(3\pi)}e^+\nu_e}$ &$\epsilon_{{\rm DT}, \eta_{(2\gamma)}\mu^+\nu_\mu}$&$\epsilon_{\eta_{(2\gamma)} \mu^+\nu_\mu}$&$\epsilon_{{\rm DT},\eta_{(3\pi)} \mu^+\nu_\mu}$&$\epsilon_{\eta_{(3\pi)} \mu^+\nu_\mu}$\\ \hline
			$K^+\pi^-\pi^-$                  &$20.39\pm0.05$&$39.93\pm0.11$  &$10.04\pm0.04$&$19.66\pm0.09$  &$18.80\pm0.05$&$36.81\pm0.11$ &$9.98\pm0.04$&$19.53\pm0.09$\\
			$K^0_S\pi^-$                     &$21.20\pm0.06$&$41.24\pm0.11$  &$10.37\pm0.04$&$20.16\pm0.09$  &$20.31\pm0.05$&$39.49\pm0.11$ &$10.50\pm0.05$&$20.42\pm0.09$\\
			$K^+\pi^-\pi^-\pi^0$             &$9.36\pm0.05$&$38.15\pm0.22$   &$4.15\pm0.04$&$16.93\pm0.17$   &$8.48\pm0.05$&$34.58\pm0.22$  &$4.03\pm0.04$&$16.44\pm0.17$\\
			$K^0_S\pi^-\pi^0$                &$10.72\pm0.05$&$40.53\pm0.21$  &$4.73\pm0.04$&$17.87\pm0.16$   &$10.41\pm0.05$&$39.37\pm0.21$ &$4.83\pm0.04$&$18.24\pm0.16$\\
			$K^0_S\pi^-\pi^-\pi^+$           &$11.07\pm0.05$&$37.31\pm0.18$  &$4.99\pm0.04$&$16.82\pm0.14$   &$10.17\pm0.05$&$34.26\pm0.18$ &$4.93\pm0.04$&$16.61\pm0.14$\\
			$K^+K^-\pi^-$                    &$15.45\pm0.05$&$37.77\pm0.13$  &$7.75\pm0.04$&$18.94\pm0.11$   &$12.79\pm0.05$&$31.28\pm0.13$ &$7.19\pm0.04$&$17.57\pm0.10$\\
			\hline
			Average& &$39.51\pm0.07$& &$18.76\pm0.06$& &$36.52\pm0.07$&&$18.60\pm0.06$\\
			\hline \hline
		\end{tabular}
	}
\end{table}

Table~\ref{tab:bf} shows the DT efficiencies ($\epsilon_{\rm DT,\eta\ell^+\nu_\ell}$) and signal efficiencies ($\epsilon_{\eta\ell^+\nu_\ell}$). A simultaneous fit is performed with a shared BF, and the resulting BFs for $\etaenu$ and $\etamunu$ are determined using Eq.~(\ref{eq:bf}).
The obtained results are summarized in Table~\ref{bf_etalnu}. The systematic uncertainties in the BF measurements are discussed below.

\begin{table*}[htp]
	\centering
	\caption{
		The signal yields in data ($N_{\rm sig}$),
		the weighted signal efficiency of detecting the signal decay in the presence of the ST $D^-$ ($\bar{\epsilon}_{\rm sig}$).
		and the obtained BFs for different signal decays ($\mathcal B_{\rm sig}$).
		For $\mathcal B_{\rm sig}$, the first uncertainties are statistical and the second are systematic.
		For other quantities, the uncertainties are statistical.
		\label{bf_etalnu}
	}
	\begin{tabular}{lccc}
		\hline\hline
		Decay&$N_{\rm sig}$&$\bar{\epsilon}_{{\rm sig}}~(\%)$&$\mathcal B_{\rm sig}~(10^{-4})$\\ \hline
		$D^+\to\eta_{(2\gamma)}e^+\nu_e$     &$1567\pm119$  & $39.51\pm0.07$  & \multirow{2}{*}{$9.75\pm0.29\pm0.28$} \\
		$D^+\to\eta_{(3\pi)}e^+\nu_e$        &$426\pm57$  & $18.76\pm0.06$  &  \\ \hline
		$D^+\to\eta_{(2\gamma)}\mu^+\nu_\mu$ &$1375\pm136$  & $36.52\pm0.07$  & \multirow{2}{*}{$9.08\pm0.35\pm0.29$} \\
		$D^+\to\eta_{(3\pi)}\mu^+\nu_\mu$    &$400\pm69$   & $18.60\pm0.06$  &  \\\hline\hline
	\end{tabular}
\end{table*}

\subsection{Systematic Uncertainties in Branching Fractions}

Table~\ref{table:bf_systot} summarizes the sources of the systematic uncertainties in the BF measurements.
They are assigned relative to the measured BFs and are discussed below. 
In the table, the systematic uncertainties of $N^{\rm tot}_{\rm ST}$, $\pi^\pm$ tracking, $\pi^\pm$ PID, $\pi^0$ and $\eta$ reconstruction, Quoted BFs,  and $E^{\rm max}_{\rm extra \gamma}$, $N_{\rm extra}^{\rm char}$\&$N_{\rm extra}^{\pi^0}$ are considered as correlated for $D^+\to\eta\mu^+\nu_\mu$ and $D^+\to\eta e^+\nu_e$. Other systematic uncertainties are independent for $D^+\to\eta\mu^+\nu_\mu$ and $D^+\to\eta e^+\nu_e$.

\paragraph{ST $D^-$ yields}

The systematic uncertainty of the fits to the $M_{\rm BC}$ spectra is estimated by varying the signal and background shapes and repeating the fits for both data and inclusive MC sample. The uncertainty due to the signal shape is evaluated by changing the matching requirement between generated and reconstructed angles from 15$^\circ$ to 10$^\circ$ or 20$^\circ$.
The uncertainty related to the background shape is obtained by shifting the endpoint by $\pm 0.2$~MeV.
In addition, the effect of removing the $M_{\rm BC}$ requirements from the ST selection is evaluated and the difference with the nominal measurement is taken as a systematic uncertainty accounting for possible mismodelling of the $M_{\rm BC}$ distribution in simulation.
These three souces of uncertainty are combined in quadrature, yielding a total systematic uncertainty of 0.3\%. 

\paragraph{$\pi^\pm$ tracking and PID}

The $\pi^\pm$ tracking and PID efficiencies are studied by using the control sample of hadronic $D\bar D$ events from $D^0\to K^-\pi^+$, $D^0\to K^-\pi^+\pi^0$, $D^0\to K^-\pi^+\pi^+\pi^-$, and $D^+\to K^-\pi^+\pi^+$~\cite{bam245}.
The systematic uncertainties in tracking and PID efficiencies of $\pi^\pm$ are assigned as 0.5\% and 0.5\% per pion, respectively.

\paragraph{$\ell^+$ tracking and PID}
\label{sec:electron}

The $\ell^+$ tracking and PID efficiencies are studied by using the control samples of $e^+e^-\to \gamma e^+e^-$ and
$e^+e^-\to \gamma \mu^+\mu^-$, respectively.
The ratios of efficiencies between the data and MC are $0.999\pm0.002$ for $e^+$ tracking and $0.969\pm0.003$ for $e^+$ PID of $D^+\to\eta e^+\nu_e$.
For $D^+\to\eta\mu^+\nu_\mu$, the corresponding ratios are $1.001\pm0.002$ for $\mu^+$ tracking and $0.986\pm0.003$ for $\mu^+$ PID.
The signal efficiencies are corrected by these factors. After correction, the uncertainties of ratios are assigned as the systematic uncertainties.

\paragraph{$\pi^0$ and $\eta$ reconstructions}
\label{sec:gamma}

The uncertainty due to the $\pi^0$ reconstruction is estimated with a control sample of $D^0\to K^-\pi^+\pi^0$, with $\bar{D}^0$ meson tagged by $K^+\pi^-$, $K^+\pi^-\pi^0$ and $K^+\pi^-\pi^-\pi^+$~\cite{pi0sys}. The systematic uncertainty in $\pi^0$ reconstruction is assigned to be 2.0\%, and that for $\eta$ reconstruction is assigned by referring to $\pi^0$ reconstruction. 

\paragraph{$M_{\eta\mu^+}$ requirement}
\label{m}

The systematic uncertainties due to the $M_{\eta\mu^+}$ requirement are estimated by changing
the nominal ranges of  [1.672,1.772] ${\rm GeV}/c^2$ and [1.660,1.760] ${\rm GeV}/c^2$
for $D^+\to\eta_{(2\gamma)}\mu^+\nu_\mu$ and $D^+\to\eta_{(3\pi)}\mu^+\nu_\mu$, respectively,
with step of 0.001 ${\rm GeV}/c^2$.
The resulting BFs are fitted with a linear function. The deviations from the nominal BFs, 0.5\% and 0.2\%, are taken as the systematic uncertainties for $D^+\to\eta_{(2\gamma)}\mu^+\nu_\mu$ and $D^+\to\eta_{(3\pi)}\mu^+\nu_\mu$, respectively.

\paragraph{$E_{\rm extra~\gamma}^{\rm max}$, $N_{\rm extra}^{\rm char}$ and $N_{\rm extra}^{\pi^0}$ requirements}
\label{em}

The systematic uncertainty in the $E_{\rm extra~\gamma}^{\rm max}$, $N_{\rm extra}^{\rm char}$ and $N_{\rm extra}^{\pi^0}$ requirements is estimated with a hadronic DT sample, which are fully reconstructed through $\bar{D}^{0}\to K^{+}\pi^{-}$ or $K^{+}\pi^{-}\pi^{-}\pi^{+}$ or $K^{+}\pi^{-}\pi^{0}$ vs. $D^{0}\to K^{-}\pi^{+}$ or $K^{-}\pi^{+}\pi^{+}\pi^{-}$ or $K^{-}\pi^{+}\pi^{0}$ and $D^{-}\to K^{+}\pi^{-}\pi^{-}$ or $K^{+}\pi^{-}\pi^{-}\pi^{0}$ or $K_{S}^{0}\pi^{-}$ or $K_{S}^{0}\pi^{-}\pi^{0}$ or $K_{S}^{0}\pi^{-}\pi^{-}\pi^{+}$ or $K^{+}K^{-}\pi^{-}$ vs. $D^{+}\to K^{-}\pi^{+}\pi^{+}$ or $K^{-}\pi^{+}\pi^{+}\pi^{0}$ or $K_{S}^{0}\pi^{+}$ or $K_{S}^{0}\pi^{+}\pi^{0}$ or $K_{S}^{0}\pi^{+}\pi^{+}\pi^{-}$ or $K^{+}K^{-}\pi^{+}$, the selection criteria are the same as that in the single tag side.
The difference in the acceptance efficiencies between data and MC simulation is assigned as the systematic uncertainty.

\paragraph{Final-state radiation recovery}

The systematic uncertainty due to final-state radiation (FSR) recovery is estimated as the difference between the results obtained with and without FSR recovery. It is assigned as 0.5\% for $D^{+}\to \eta e^+\nu_{e}$.

\paragraph{$U_{\rm miss}$ fit}

The systematic uncertainty due to the $U_{\rm miss}$ fit is considered in two parts.
Since a Gaussian function is convolved with the simulated signal shapes to account for the resolution difference between
data and MC simulation, the systematic uncertainty from the signal shape is ignored.
The systematic uncertainty due to background shapes is considered for both peaking and combinatorial backgrounds.
The effect of the fixed peaking backgrounds is examined by varying the fixed yields of the peaking backgrounds by $\pm1\sigma$. The uncertainty comes from the uncertainty of its branching fraction, and we take the difference between the new branching fraction and the normal value as the systematic uncertainty. 
The effect of the combinatorial background shapes is estimated by varying the yields of dominant background channels within uncertainty of their input BFs in the inclusive MC sample.
The changes in the BFs are taken as  the corresponding systematic uncertainties.

\paragraph{MC statistics}

The relative uncertainties in the signal efficiencies are assigned as a source of systematic uncertainty due to the limited size of the MC sample.

\paragraph{MC model}

The detection efficiencies are determined using signal MC samples generated with the hadronic transition form factors measured in this analysis.
The corresponding systematic uncertainties are estimated by varying the parameters by $\pm1\sigma$.

\paragraph{Quoted BFs}

For the $D^+\to\eta_{(2\gamma)}\ell^+\nu_\ell$ decays,
the uncertainty of the quoted BF of $\eta\to\gamma\gamma$ is 0.5\%~\cite{pdg2022}.
For the $D^+\to\eta_{(3\pi)}\ell^+\nu_\ell$ decays,
the combined uncertainty in the quoted BFs of $\eta\to\pi^+\pi^-\pi^0$ and $\pi^0\to\gamma\gamma$ is 1.1\% .

\begin{table}[htbp]
	\caption{Relative systematic uncertainties (in unit of \%) in the measurements of the BFs, where ``...'' denotes the unapplicable items for the experimental measurement.
		\label{table:bf_systot}}
	\centering
	\resizebox{1.0\textwidth}{!}{
		\begin{tabular}{lcccc}
			\hline
			\hline

			Source                              & $D^+\to\eta_{(2\gamma)}e^+\nu_e$ & $D^+\to\eta_{(3\pi)}e^+\nu_e$ &$D^+\to\eta_{(2\gamma)}\mu^+\nu_\mu$ & $D^+\to\eta_{(3\pi)}\mu^+\nu_\mu$\\\hline
			$N^{\rm tot}_{\rm ST}$                &0.3    &0.3     &0.3     &0.3    \\
			$\pi^\pm$ tracking                    &...    &1.0     &...     &1.0    \\
			$\pi^\pm$ PID                         &...    &1.0     &...     &1.0    \\
			$\pi^0$ and $\eta$ reconstruction     &2.0    &2.0     &2.0     &2.0    \\
			$E^{\rm max}_{\rm extra \gamma}$, $N_{\rm extra}^{\rm char}$\&$N_{\rm extra}^{\pi^0}$
			&0.1    &0.1     &0.1     &0.1    \\
			Quoted BFs            &0.5    &1.1     &0.5     &1.1    \\
			\hline
			$\ell^+$ tracking                     &0.2    &0.2     &0.2     &0.2    \\
			$\ell^+$ PID                          &0.3    &0.3     &0.3     &0.3    \\
			$M_{\eta\mu}$                         &...    &...     &0.5     &0.2    \\
			FSR recovery                           &0.5  &0.5   &...      &...\\
			$U_{\rm miss}$ fit                    &0.2    &0.2     &0.4     &0.3    \\
			MC  statistics                        &0.2    &0.3     &0.2     &0.3    \\
			MC model                              &1.8    &1.8     &0.8     &0.8    \\
			\hline
			Total                                 &2.8    &3.3       &2.4   &2.9 \\
			\hline\hline
		\end{tabular}
			}
\end{table}

\section{HADRONIC TRANSITION FORM FACTORS}

\subsection{Theoretical Formula}

In theory, the partial   decay width of the semileptonic decay $D^+ \to \eta \ell^+\nu_\ell$ can be expressed as
\begin{equation}
	\frac{{\rm d}\Gamma\left(D^+ \to \eta\ell^+\nu_\ell \right)}{{\rm d}q^2} = \frac{G_F^2|V_{cd}|^2}{24\pi^3}\left |f^\eta_+(q^2)\right |^2\left |\vec p_{\eta}\right |^3,
\end{equation}
where $q$ is the momentum transfer of the $\ell^+\nu_\ell$ system, $|\vec p_\eta|$ is the magnitude of the $\eta$ 3-momentum in the $D^+$ rest frame and $G_F$ is the Fermi coupling constant.
The two-parameter series expansion [$z$ series (2 par.)]~\cite{Becher:2005bg} is a widely used model to describe
the hadronic transition form factor of semileptonic $D^+$ decays. It is given by
	\begin{equation}
		\begin{array}{l}
			f^\eta_+(q^2)=\frac{1}{P(q^2)\Phi(q^2)}\frac{f^\eta_+(0)P(0)\Phi(0)}{1+r_{1}(t_0)z(0,t_0)}
			\times\left(1+r_{1}(t_0)[z(q^2,t_0)]\right).
		\end{array}
	\end{equation}
Here, $t_0=t_+(1-\sqrt{1-t_{-}/t_+})$, $t_{\pm}=(m_{D^+}\pm m_\eta)^2$, $m_{D^+}$ and $m_\eta$ are the masses of $D^+$ and $\eta$ particles, $P(q^2)=z(q^2,m_{D^{*}}^2)$, $m_{D^*}$ is the pole mass of the vector form factor accounting for the strong interaction between
$D^+$ and $\eta$ mesons and usually taken as the mass of the lowest lying $c\bar d$ vector meson $D^*$~\cite{pdg2022}, $z(q^2,t_0)=\frac{\sqrt{t_+-q^2}-\sqrt{t_+-t_0}}{\sqrt{t_+-q^2}+\sqrt{t_+-t_0}}$, and $r_1$ is a free parameter. $\Phi$ is given by
	\begin{equation}
		\begin{array}{l}
			\Phi(q^2)=\sqrt{\frac{1}{24\pi\chi_{V}}}\left(\frac{t_+-q^2}{t_+-t_0}\right)^{1/4}\left(\sqrt{t_+-q^2}+\sqrt{t_+}\right)^{-5}\times\\
			\left(\sqrt{t_+-q^2}+\sqrt{t_+-t_0}\right)\left(\sqrt{t_+-q^2}+\sqrt{t_+-t_{-}}\right)^{3/2}
			\times\left(t_+-q^2\right)^{3/4},
		\end{array}
	\end{equation}
where $\chi_{V}$ is obtained from dispersion
relations using perturbative QCD~\cite{chiV}.

\subsection{Partial Decay Rates}

To extract the hadronic transition form factors for semileptonic $D^+$ decays, the partial decay rates are measured in four $q^2$ ranges ($N_{\rm bins}$): $(m_\ell^2,0.3)$, $(0.3,0.7)$, $(0.7,1.3)$, and $(1.3,1.8)$~GeV$^2/c^4$, where $m_\ell$ is the mass of $\ell$. The partial decay rate in the $i$-th $q^2$ interval is determined as
\begin{equation}
	\frac{{\rm d}\Gamma_{i}}{{\rm d}q_{i}^2}=\frac{\Delta\Gamma^{\rm measured}_{i}}{\Delta q^2_{i}},
\end{equation}
where
$\Delta\Gamma^{\rm measured}_{i}=\frac{N_{\rm produced}^{i}}{\tau_{D^+}\cdot N_{\rm tag}}$
is the partial decay rate in the $i$-th $q^2$ interval, $N_{\rm produced}^{i}$ is the number of events produced in the $i$-th $q^2$ interval,
$\tau_{D^+}$ is the $D^+$ lifetime~\cite{pdg2022} and $N_{\rm{tag}}$ is the number of the ST $D^-$ mesons.

In the $i$-th $q^2$ interval, the number of events produced in data is calculated as
\begin{equation}
	N_{\rm produced}^{i}=\sum_j^{N_{\rm bins}}\left(\epsilon^{-1}\right)_{ij}N_{\rm DT}^{j},
\end{equation}
where $(\epsilon^{-1})_{ij}$ is the element of the inverse signal efficiency matrix for events produced in the $j$-th $q^2$ interval that are reconstructed in the $i$-th $q^2$ interval, obtained by analyzing the signal MC events.
The statistical uncertainty of $N_{\rm produced}^{i}$ is given by
\begin{equation}
	\left[\sigma\left(N_{\rm produced}^{i}\right)\right]^2=\sum_j^{N_{\rm bins}}\left(\epsilon^{-1}\right)_{ij}^2\left[\sigma\left(N_{\rm DT}^{j}\right)\right]^2,
\end{equation}
where $\sigma(N_{\rm DT}^{j})$ is the statistical uncertainty of $N_{\rm DT}^{j}$.
The signal efficiency matrix $\epsilon_{ij}$ is given by
\begin{equation}
	\epsilon_{ij}=\frac{N_{ij}^{\rm reconstructed}}{N_j^{\rm generated}}\cdot \frac{1}{\epsilon_{\rm tag}},
\end{equation}
where $N_{ij}^{\rm reconstructed}$ is the number of events generated in the $j\text{-}$th $q^2$ interval that are reconstructed in the $i$-th $q^2$ interval, $N_j^{\rm generated}$ is the total number of events generated in the $j\text{-}$th $q^2$ interval, and $\epsilon_{\rm tag}$ is the efficiency of detecting ST $D^-$ mesons.

Tables~\ref{etaenu_effmatrix} and ~\ref{etamunu_effmatrix} give the elements of the signal efficiency matrices weighted by the yields of each ST $D^-$ meson MODE in data.

\begin{table}[htbp]
	\caption{The weighted efficiency matrix (in unit of \%) for $D^+\to\eta e^+\nu_e$.
		\label{etaenu_effmatrix}}
	\centering
			\begin{tabular}{c|cccc|cccc}
				\hline\hline
				& \multicolumn{4}{c|}{$D^+\to\eta_{(2\gamma)}e^+\nu_e$ }& \multicolumn{4}{c}{$D^+\to\eta_{(3\pi)}e^+\nu_e$ } \\
				$q^2$ bin &1&2&3&4&1&2&3&4\\
				\hline
				1&36.03& 1.20 &  0.00&  0.00 &18.22& 0.43&  0.00&  0.00 \\
				2&1.35&  37.04& 1.17&  0.00  &0.75&  17.88& 0.37&  0.00 \\
				3&0.05&  1.33&  39.58& 1.91  &0.03&  0.80&  17.57& 0.61 \\
				4&0.01&  0.03&  0.56&  40.81 &0.00&  0.01&  0.37&  16.76\\
				\hline\hline
			\end{tabular}
	\end{table}

	\begin{table}[htbp]
		\caption{The weighted efficiency matrix (in unit of \%) for $D^+\to\eta \mu^+\nu_\mu$.
			\label{etamunu_effmatrix}}
		\centering
			\begin{tabular}{c|cccc|cccc}
				\hline\hline
				& \multicolumn{4}{c|}{$D^+\to\eta_{(2\gamma)}\mu^+\nu_\mu$ }& \multicolumn{4}{c}{$D^+\to\eta_{(3\pi)}\mu^+\nu_\mu$ } \\
				$q^2$ bin &1&2&3&4&1&2&3&4\\
				\hline
				1&33.21& 0.89&  0.00&  0.00 &17.90& 0.29&  0.00&  0.01\\
				2&1.34&  35.12& 0.97&  0.00 &0.72&  18.20& 0.33&  0.00\\
				3&0.04&  1.07&  36.35& 1.47 &0.03&  0.72&  17.57& 0.53\\
				4&0.01&  0.02&  0.40&  36.23&0.01&  0.01&  0.26&  16.14\\
				\hline\hline
			\end{tabular}
	\end{table}

\begin{figure}[htbp]
		\begin{center}
			\subfigure{\includegraphics[width=1.0\textwidth]{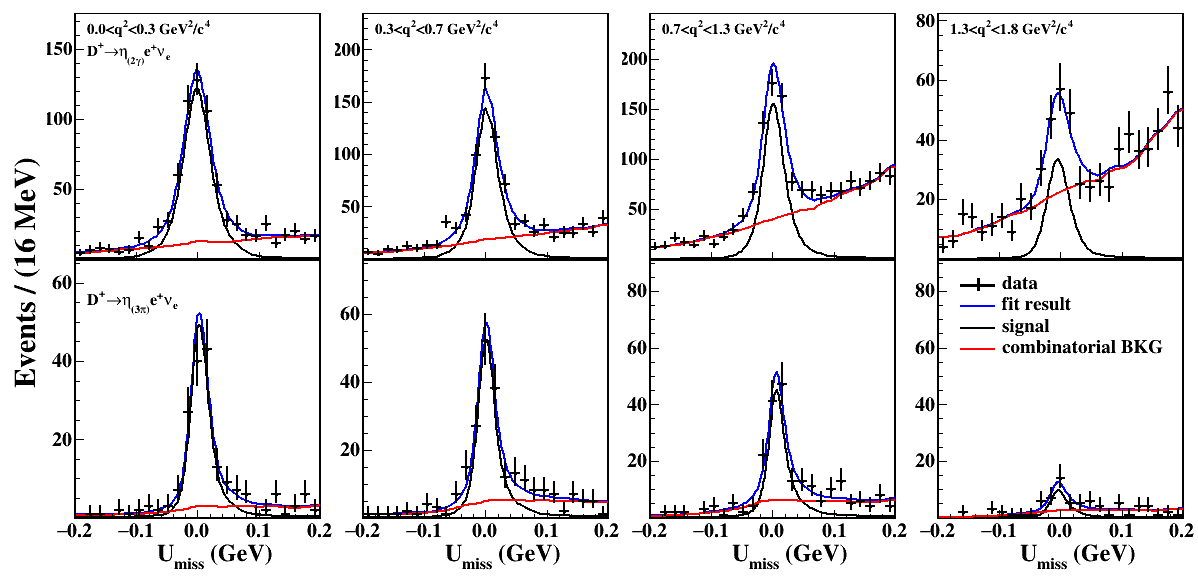}}
			\caption{
				Fits to the $U_{\rm miss}$ distributions of the candidates for the semileptonic decay $D^+\to\eta e^+\nu_e$ in different $q^2$  bins.
				The dots with error bars are data. The blue solid curves are the fit results,
				the black solid curves are the signal shapes, and
				the black dashed curves are the fitted combinatorial background shapes.
				\label{etaenua_umissq2}
			}
		\end{center}
	\end{figure}

\begin{figure}[htbp]
		\begin{center}
			\subfigure{\includegraphics[width=1.0\textwidth]{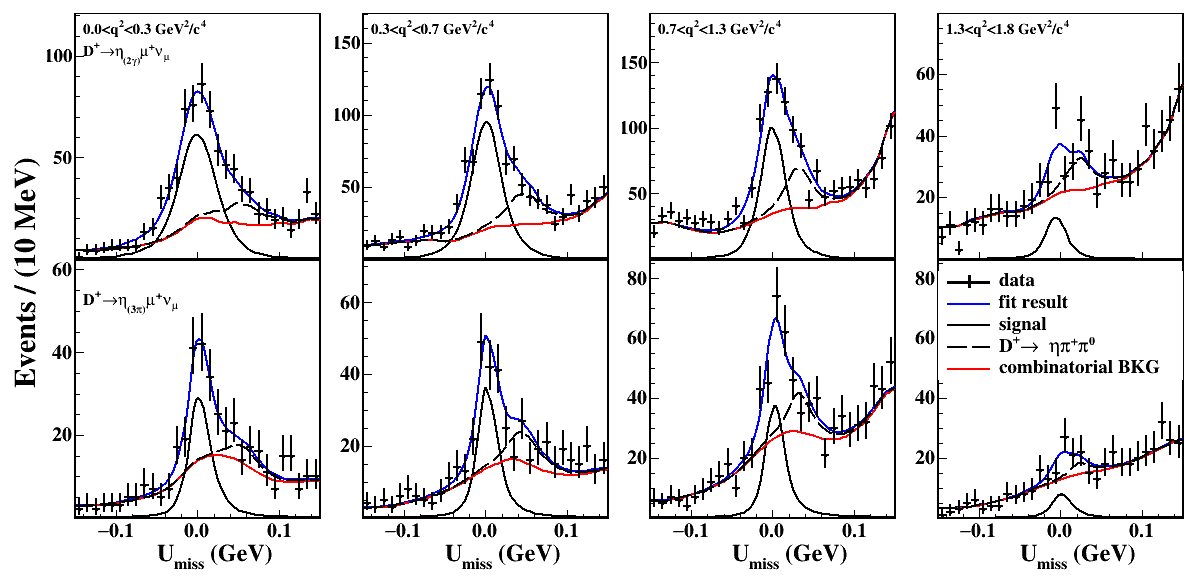}}
			\caption{
				Fits to the $U_{\rm miss}$ distributions of the candidates for the semileptonic decay $D^+\to\eta\mu^+\nu_\mu$ in different $q^2$ bins.
				The dots with  error bars are data. The blue solid curves are the fit results,
				the black solid curves are the signal shapes,
				the black dashed curves are the peaking backgrounds stacked to the combinatorial background, and
				the red dashed curves are the fitted combinatorial background shapes.
				\label{etamunua_umissq2}
			}
		\end{center}
	\end{figure}

	For each signal decay, the signal yield observed in each reconstructed $q^2$ interval is obtained
	from a fit to the $U_{\rm miss}$ distribution.
	Figures~\ref{etaenua_umissq2} and~\ref{etamunua_umissq2} show the results of the fits to the $U_{\rm miss}$ distributions in the reconstructed $q^2$ bins.

	Tables~\ref{tab:etaenua_decayrate} and \ref{tab:etamunua_decayrate}  summarize the $q^2$ ranges, the fitted numbers of observed DT events ($N^{i}_{\rm DT}$),
	the numbers of produced events ($N^{i}_{\rm produced}$) calculated by the weighted efficiency matrix
	and the partial decay rates of semileptonic $D^+$ decays ($\Delta\Gamma^{\rm measured}_{i}$) in the individual $q^2$ bins.

	\begin{table}[htbp]
		\caption{ The measured partial decay rates of the semileptonic decay $D^+\to\eta_{(2\gamma)}e^+\nu_e$ in various $q^2$ bins, where
			$N^{i}_{\rm DT}$ is the observed DT yield, $N^{i}_{\rm produced}$ is the produced yield,
			$\Delta\Gamma^{\rm measured}_{i}$ is the partial decay rate, and the uncertainties are statistical. \label{tab:etaenua_decayrate}}
		\centering
		\resizebox{1.0\textwidth}{!}{
		\begin{tabular}{c|ccc|ccc}
			\hline\hline
			& \multicolumn{3}{c|}{$D^+\to\eta_{(2\gamma)}e^+\nu_e$ }& \multicolumn{3}{c}{$D^+\to\eta_{(3\pi)}e^+\nu_e$ } \\
			$q^2~({\rm GeV^{2}}/c^4)$&  $N^{i}_{\rm DT}$&     $N^{i}_{\rm produced}$&     $\Delta\Gamma^{\rm measured}_{i}~({\rm ns}^{-1})$ &  $N^{i}_{\rm DT}$&     $N^{i}_{\rm produced}$&     $\Delta\Gamma^{\rm measured}_{i}~({\rm ns}^{-1})$\\
			\hline
			$(m_e^2,0.3)$&   $485\pm27 $&  $3316\pm188$& $0.301\pm0.017  $&$137\pm14 $&  $3265\pm339$& $0.296\pm0.031 $\\
			$(0.3,0.7)$&   $489\pm28 $&  $3137\pm191$& $0.284\pm0.017 $ &$146\pm16 $&  $3440\pm388$& $0.312\pm0.035 $\\
			$(0.7,1.3)$&   $492\pm31$&   $3024\pm198$& $0.274\pm0.018  $&$124\pm15$&   $2953\pm380$& $0.268\pm0.034$\\
			$(1.3,1.8)$&   $100\pm16$&   $579\pm99$&   $0.052\pm0.009 $ &$28\pm8$&     $667\pm200$&  $0.060\pm0.018  $\\
			\hline\hline
		\end{tabular}
	}
	\end{table}

	\begin{table}[htbp]
		\caption{ The measured partial decay rates of the semileptonic decay $D^+\to\eta\mu^+\nu_\mu$ in different $q^2$ bins, where
			$N^{i}_{\rm DT}$ is the observed DT yield, $N^{i}_{\rm produced}$ is the produced yield,
			$\Delta\Gamma^{\rm measured}_{i}$ is the partial decay rate, and the uncertainties are statistical. \label{tab:etamunua_decayrate}}
		\centering
		\resizebox{1.0\textwidth}{!}{
		\begin{tabular}{c|ccc|ccc}
			\hline\hline
			& \multicolumn{3}{c|}{$D^+\to\eta_{(2\gamma)}\mu^+\nu_\mu$ }& \multicolumn{3}{c}{$D^+\to\eta_{(3\pi)}\mu^+\nu_\mu$ } \\
			$q^2~({\rm GeV^{2}}/c^4)$&   $N^{i}_{\rm DT}$&  $N^{i}_{\rm produced}$&   $\Delta\Gamma^{\rm measured}_{i}~({\rm ns}^{-1})$&   $N^{i}_{\rm DT}$&  $N^{i}_{\rm produced}$&   $\Delta\Gamma^{\rm measured}_{i}~({\rm ns}^{-1})$\\
			\hline
			$(m_\mu^2,0.3)$&   $400\pm33 $&  $2968\pm254$& $0.269\pm0.023 $  & $115\pm18 $&  $2822\pm459$& $0.256\pm0.042  $\\
			$(0.3,0.7)$&   $504\pm31 $&  $3447\pm223$& $0.313\pm0.020 $  & $127\pm16 $&  $2942\pm382$& $0.267\pm0.035 $ \\
			$(0.7,1.3)$&   $442\pm31 $&  $2966\pm217$& $0.269\pm0.020  $ & $135\pm21 $&  $3277\pm532$& $0.297\pm0.048  $\\
			$(1.3,1.8)$&   $50\pm15 $&   $448\pm47$&   $0.041\pm0.004  $ & $30\pm13 $&   $782\pm349$&  $0.071\pm0.032  $\\
			\hline\hline
		\end{tabular}
	}
	\end{table}

\begin{figure}[htbp]
	\begin{center}
		\includegraphics[width=0.72\textwidth]{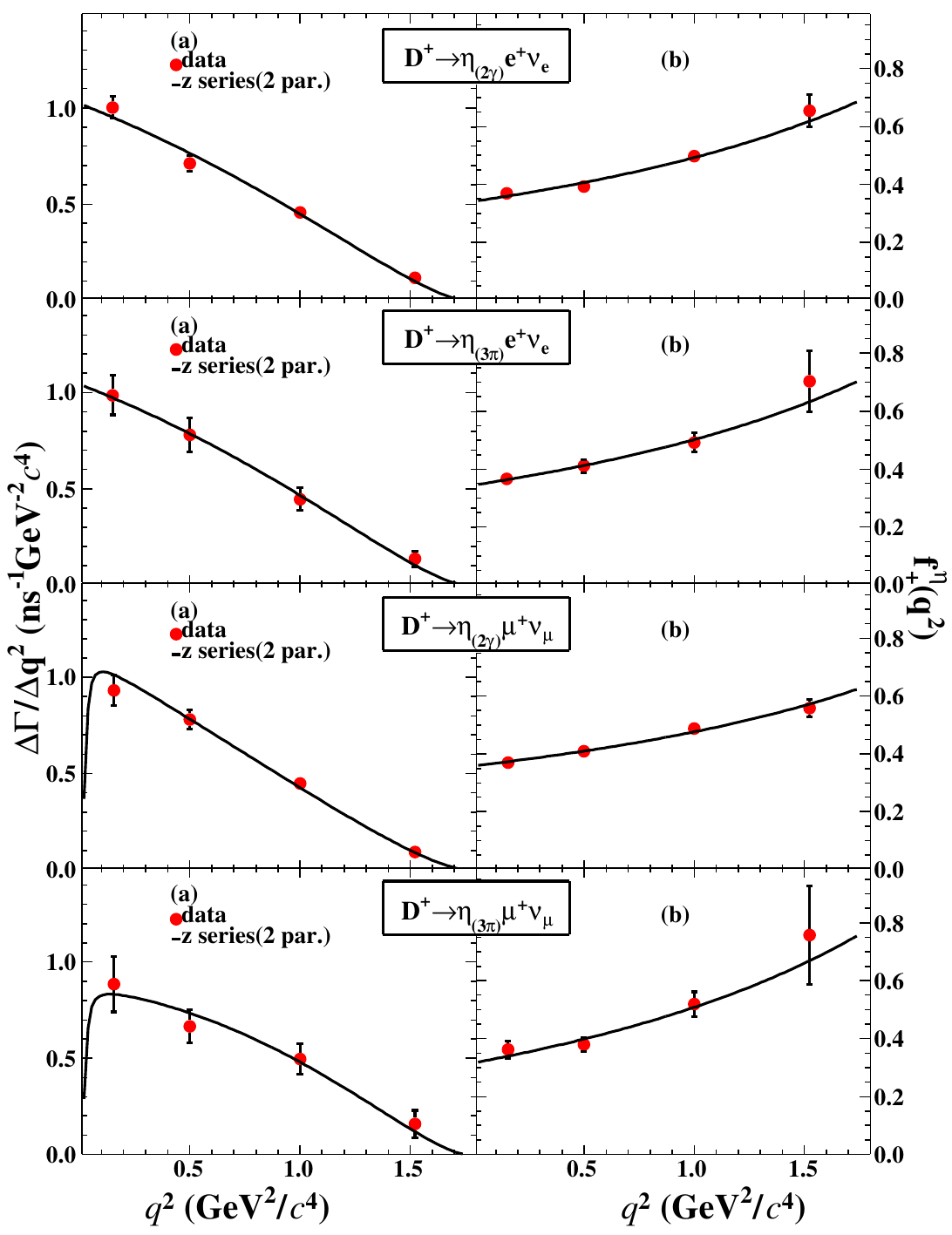}
		\caption{(Left) Separate fits to the partial decay rates of the four semileptonic decays $D^+\to\eta\ell^+\nu_\ell$ and (Right) projections of the hadronic form factor as a function of $q^2$. The dots with error bars are the measured partial decay rates and the solid curves are the fits.
			\label{fig:ff_etalnu}
		}
	\end{center}
\end{figure}

\subsection{Construction of $\chi^2$ and Statistical Covariance Matrices}

	To extract the hadronic transition form factor parameters and $|V_{cd}|$,
	the least $\chi^2$ method is used to fit the partial decay rates of the semileptonic $D^+$ decays.
	Considering the correlations among the measured partial decay rates $\Delta\Gamma_i^{\rm measured}$ in different $q^2$ bins, the $\chi^2$ is defined as
		\begin{equation*}
			\chi^2 = \sum_{i,j=1}^{N_{\rm bins}}\left(\Delta\Gamma_i^{\rm measured}-\Delta\Gamma_i^{\rm expected}\right) (C^{-1})_{ij}\left(\Delta\Gamma_j^{\rm measured}-\Delta\Gamma_j^{\rm expected}\right),
			\label{eq:chi}
		\end{equation*}
	where $\Delta\Gamma_i^{\rm expected}$ is the expected decay rate in the $i$-th channel, $C_{ij}$ is the element of the covariance matrix of the measured partial decay rates and it is given by $C_{ij} = C_{ij}^{\mathrm{stat}}+C_{ij}^{\rm syst}$.
	Here, $C_{ij}^{\rm stat}$ and $C_{ij}^{\rm syst}$ are elements of the
	statistical and systematic covariance matrices, respectively.
	The elements of the statistical covariance matrix are defined as
	\begin{equation}
		C_{ij}^{\rm stat} =\left (\frac{1}{\tau_{D^{+}_s}N_{\rm tag}}\right)^2\sum_{\alpha}\epsilon_{i\alpha}^{-1}\epsilon_{j\alpha}^{-1}\left(\sigma\left(N_{\rm DT}^{\alpha}\right)\right)^2.
	\end{equation}
	where $\sigma(N_{\rm DT}^{\alpha})$ is the statistical uncertainty of the signal yield observed in the $\alpha$-th interval.
	Tables~\ref{tab:etaenua_statmatrix} and \ref{tab:etamunua_statmatrix}  give the elements of the statistical covariance matrices for $D^+\to \eta e^+\nu_e$ and $D^+\to \eta \mu^+\nu_\mu$, respectively.

	\begin{table}[htbp]
		\caption{The statistical covariance matrices for $D^+\to\eta e^+\nu_e$,
			where $i$ and $j$ denote the produced and reconstructed bins, respectively.
			\label{tab:etaenua_statmatrix}}
		\centering
		\begin{tabular}{c|cccc|cccc}
			\hline\hline
			& \multicolumn{4}{c|}{$D^+\to\eta_{(2\gamma)}e^+\nu_e$ }& \multicolumn{4}{c}{$D^+\to\eta_{(3\pi)} e^+\nu_e$ } \\
			$\epsilon_{ij}$  &1&2&3&4 &1&2&3&4\\
			\hline
			1	&1.000  &-0.070 &0.002  &-0.000	&1.000  &-0.064 &0.002  &-0.000	\\
			2	&-0.070 &1.000  &-0.065 &0.001	&-0.064 &1.000  &-0.067 &0.002  \\
			3   &0.002  &-0.065 &1.000  &-0.051 &0.002  &-0.067 &1.000  &-0.060  \\
			4   &-0.000 &0.001  &-0.051 &1.000  &-0.000 &0.002  &-0.060 &1.000 \\
			\hline
			\hline
		\end{tabular}
	\end{table}

	\begin{table}[htbp]
		\caption{The statistical covariance matrices for $D^+\to\eta\mu^+\nu_\mu$,
			where $i$ and $j$ denote the produced and reconstructed bins, respectively.
			\label{tab:etamunua_statmatrix}}
		\centering
		\begin{tabular}{c|cccc|cccc}
			\hline\hline
			& \multicolumn{4}{c|}{$D^+\to\eta_{(2\gamma)}\mu^+\nu_\mu$ }& \multicolumn{4}{c}{$D^+\to\eta_{(3\pi)}\mu^+\nu_\mu$ } \\
			$\epsilon_{ij}$  &1&2&3&4  &1&2&3&4\\
			\hline
			1	&1.000  &-0.067 &0.001  &-0.001 &1.000  &-0.061 &0.001  &-0.001\\
			2	&-0.067 &1.000  &-0.057 &0.001	&-0.061 &1.000  &-0.055 &0.001 \\
			3   &0.001  &-0.057 &1.000  &-0.059 &0.001  &-0.055 &1.000  &-0.044 \\
			4   &-0.001 &0.001  &-0.059 &1.000  &-0.001 &0.001  &-0.044 &1.000\\
			\hline
			\hline
		\end{tabular}
	\end{table}

	\subsection{Systematic Uncertainties in Hadronic Form Factor}

	Several sources of systematic uncertainties are discussed below.

	\paragraph{Number of $D^{-}$ tags}

	The uncertainties associated with the number of $D^-$ tags are fully correlated across $q^{2}$ bins. The systematic covariance matrix is calculated by
	\begin{equation}
		C_{ij}^{\mathrm{sys}}(N_{\mathrm{tag}})=\Delta\Gamma_{i}\Delta\Gamma_{j}(\frac{\sigma(N_{\mathrm{tag}})}{N_{\mathrm{tag}}})^{2},
	\end{equation}
	where $\sigma(N_{\mathrm{tag}})/N_{\mathrm{tag}}$ is the relative uncertainty of the number of $D^-$ tags.

	\paragraph{$D^+$ lifetime}

	The systematic uncertainties associated with the $D^+$ lifetime are fully correlated across the $q^2$ bins. The element of the related systematic covariance matrix is calculated by
	\begin{equation}
		C_{ij}^{\rm syst}\left(\tau_{D^+}\right)=\sigma\left(\Delta\Gamma_i\right)\sigma\left(\Delta\Gamma_j\right),
	\end{equation}
	where $\sigma(\Delta\Gamma_i)=\sigma \tau_{D^+}\cdot\Delta\Gamma_i$ and $\sigma \tau_{D^+}$ is the uncertainty of the $D^+$ lifetime~\cite{pdg2022}.

	\paragraph{MC statistics}

	The systematic uncertainty and correlation in $q^2$ bins due to the limited size of the MC samples is calculated by
	\begin{equation}
		C_{ij}^{\rm syst}=\left(\frac{1}{\tau_{D^+}N_{\rm tag}}\right)^2\sum_{\alpha\beta}N_{\rm DT}^{\alpha}N_{\rm DT}^{\beta}\mathrm{Cov}\left(\epsilon_{i\alpha}^{-1},\epsilon_{j\beta}^{-1}\right),
	\end{equation}
	where the covariances of the inverse efficiency matrix elements are given by~\cite{Lefebvre:1999yu}
	\begin{equation}
		\mathrm{Cov}\left(\epsilon_{i\alpha}^{-1},\epsilon_{j\beta}^{-1}\right)=\sum\limits_{mn}\left(\epsilon_{im}^{-1}\epsilon_{j m}^{-1}\right)\left [\sigma\left (\epsilon_{mn}\right )\right]^2\left (\epsilon_{\alpha n}^{-1}\epsilon_{\beta n}^{-1}\right).
	\end{equation}

	\paragraph{Tracking and PID}

	The systematic uncertainties associated with $\ell^+$ or $\pi^\pm$ tracking and PID efficiencies are estimated by varying the corresponding correction factors within $\pm 1\sigma$. Using the new efficiency matrix, the element of the corresponding systematic covariance matrix is calculated by
	\begin{equation}
		C_{ij}^{\rm syst}\left (\rm{Tracking~and~PID}\right)=\delta\left(\Delta\Gamma_i\right)\delta\left(\Delta\Gamma_j\right),
	\end{equation}
	where $\delta(\Delta\Gamma_i)$ denotes the change in the partial decay rate in the $i$-th $q^2$ interval.

	\paragraph{$U_{\rm miss}$ fit}

	The systematic covariance matrix arising from the uncertainty in the $U_{\rm miss}$ fit has elements
	\begin{equation}
		C_{ij}^{\rm syst}\left(U_{\rm miss}~\rm{fit}\right)=\left(\frac{1}{\tau_{D^+}N_{\rm tag}}\right)^2\sum_{\alpha}\epsilon_{i\alpha}^{-1}\epsilon_{j\alpha}^{-1}\left(\sigma_{\alpha}^{\rm{fit}}\right)^2,
	\end{equation}
	where $\sigma_{\alpha}^{\rm{fit}}$ is the systematic uncertainty of the signal yield in the $\alpha$-th interval obtained by varying the background shape in the $U_{\rm miss}$ fit.

	\paragraph{Partial decay rates}

	Systematic uncertainties associated with the partial decay rates are estimated by comparing the difference in efficiencies between series expansion model and modified pole model. The related systematic covariance matrices are calculated by
	\begin{equation}
		C_{ij}^{\mathrm{sys}}(\mathrm{F.F.})=\delta(\Delta\Gamma_{i})\delta(\Delta\Gamma_{j}),
	\end{equation}
	where $\delta(\Delta\Gamma_{i})$ denotes the change of the partial decay rate in the $i$-th $q^{2}$ bin.

	\paragraph{Other uncertainties}

        Other systematic uncertainties, include the $E_{\rm extra ~\gamma}^{\rm max}$, $N^{\rm char}_{\rm extra}$ and $N_{\rm extra}^{\pi^0}$ requirements, $\pi^0(\eta)$ reconstruction, quoted BFs, and FSR recovery are assumed to be  fully correlated across $q^2$ bins and the element of the corresponding systematic covariance matrix is calculated by
	\begin{equation}
		C_{ij}^{\rm syst}=\sigma\left(\Delta\Gamma_i\right)\sigma\left(\Delta\Gamma_j\right),
	\end{equation}
	where $\sigma(\Delta\Gamma_i)=\sigma_{\rm syst}\cdot\Delta\Gamma_i$ and $\sigma_{\rm syst}$ is the corresponding uncertainty,
	as summarized in Tables \ref{tab:etaenua_sysq2} and \ref{tab:etamunua_sysq2} for
	$D^+\to\eta e^+\nu_e$ and $D^+\to\eta \mu^+\nu_\mu$, respectively. In the tables, the upper part corresponds to correlated uncertainties and the lower part corresponds to uncorrelated uncertainties.
	Tables~\ref{tab:etaenua_sysmatrix} and \ref{tab:etamunua_sysmatrix} provide the systematic covariance matrices for individual signal decays.

	\begin{table}[htbp]
		\caption{The systematic uncertainties (in units of \%) of the measured decay rates of $D^+\to\eta e^+\nu_e$ in different
			$q^2$ bins, where ``...'' denotes the unapplicable items for the experimental measurement.
			\label{tab:etaenua_sysq2}}
		\centering
		\begin{tabular}{c|cccc|cccc}
			\hline\hline
			& \multicolumn{4}{c|}{$D^+\to\eta_{(2\gamma)}e^+\nu_e$ }& \multicolumn{4}{c}{$D^+\to\eta_{(3\pi)} e^+\nu_e$ } \\
			$q^2$ bin &1&2&3&4&1&2&3&4\\
			\hline
			$N_{\rm ST}^{\rm tot}$                                     &0.3&0.3&0.3&0.3&0.3&0.3&0.3&0.3\\
			$D^+$ lifetime                                             &0.5&0.5&0.5&0.5&0.5&0.5&0.5&0.5\\
			$\pi^\pm$ tracking                                         &...&...&...&...&1.0&1.0&1.0&1.0\\
			$\pi^\pm$ PID                                              &...&...&...&...&1.0&1.0&1.0&1.0\\
			$\pi^0(\eta)$ reconstruction                               &2.0&2.0&2.0&2.0&2.0&2.0&2.0&2.0\\
			$E_{\rm extra,~\gamma}^{\rm max}, N^{\rm char}_{\rm extra}\&N_{\rm extra}^{\pi^0}$&0.1&0.1&0.1&0.1&0.1&0.1&0.1&0.1\\
			Quoted $\mathcal B$                                        &0.5&0.5&0.5&0.5&1.1&1.1&1.1&1.1\\
			$U_{\rm miss}$ fit                                         &0.2&0.2&0.2&0.2&0.2&0.2&0.2&0.2\\
			\hline
			$e^+$ tracking                                             &0.2&0.2&0.2&0.2&0.2&0.2&0.2&0.2\\
			$e^+$ PID                                                  &0.3&0.3&0.3&0.3&0.3&0.3&0.3&0.3\\
			FSR recovery                                             &0.5&0.5&0.5&0.5&0.5&0.5&0.5&0.5\\
			MC statistics                                              &0.3&0.2&0.3&0.6&0.4&0.4&0.4&1.0\\
			MC model                                                   &1.8&1.8&1.8&1.8&1.8&1.8&1.8&1.8\\
			\hline
			Total                                                      &2.9&2.9&2.9&2.9&3.4&3.4&3.4&3.5\\
			\hline
			\hline
		\end{tabular}
	\end{table}

	\begin{table}[htbp]
		\caption{The systematic uncertainties (in unit of \%) of the measured decay rates of $D^+\to\eta\mu^+\nu_\mu$ in different
			$q^2$ bins, where ``...'' denotes the unapplicable items for the experimental measurement.
			\label{tab:etamunua_sysq2}}
		\centering
		\begin{tabular}{c|cccc|cccc}
			\hline\hline
			& \multicolumn{4}{c|}{$D^+\to\eta_{(2\gamma)}\mu^+\nu_\mu$ }& \multicolumn{4}{c}{$D^+\to\eta_{(3\pi)}\mu^+\nu_\mu$ } \\
			$q^2$ bin &1&2&3&4 &1&2&3&4\\
			\hline
			$N_{\rm ST}^{\rm tot}$                                     &0.3&0.3&0.3&0.3&0.3&0.3&0.3&0.3\\
			$D^+$ lifetime                                             &0.5&0.5&0.5&0.5&0.5&0.5&0.5&0.5\\
			$\pi^\pm$ tracking                                         &...&...&...&...&1.0&1.0&1.0&1.0\\
			$\pi^\pm$ PID                                              &...&...&...&...&1.0&1.0&1.0&1.0\\
			$\pi^0(\eta)$ reconstruction                               &2.0&2.0&2.0&2.0&2.0&2.0&2.0&2.0\\
			$E_{\rm extra,~\gamma}^{\rm max}, N^{\rm char}_{\rm extra}\&N_{\rm extra}^{\pi^0}$&0.1&0.1&0.1&0.1&0.1&0.1&0.1&0.1\\
			Quoted $\mathcal B$                                        &0.5&0.5&0.5&0.5&1.1&1.1&1.1&1.1\\
			$U_{\rm miss}$ fit                                         &0.4&0.4&0.4&0.4&0.3&0.3&0.3&0.3\\
			\hline
			$\mu^+$ tracking                                           &0.2&0.2&0.2&0.2&0.2&0.2&0.2&0.2\\
			$\mu^+$ PID                                                &0.5&0.3&0.1&0.2&0.4&0.3&0.1&0.2\\
			$M_{\eta\mu}$                                              &0.5&0.5&0.5&0.5&0.2&0.2&0.2&0.2\\
			MC statistics                                              &0.3&0.2&0.3&0.6&0.4&0.4&0.4&0.9\\
			MC model                                                   &0.8&0.8&0.8&0.8&0.8&0.8&0.8&0.8\\
			\hline
			Total                                                      &2.5&2.4&2.4&2.5&2.9&2.9&2.9&3.0\\
			\hline
			\hline
		\end{tabular}
	\end{table}

	\begin{table}[htbp]
		\caption{The systematic correlation matrices for $D^+\to\eta e^+\nu_e$.
			\label{tab:etaenua_sysmatrix}}
		\centering
		\begin{tabular}{c|cccc|cccc}
			\hline
			\hline
			& \multicolumn{4}{c|}{$D^+\to\eta_{(2\gamma)}e^+\nu_e$ }& \multicolumn{4}{c}{$D^+\to\eta_{(3\pi)} e^+\nu_e$ } \\
			$\epsilon_{ij}$&1&2&3&4&1&2&3&4\\ \hline
			1&1.000&0.944&0.947&0.841&1.000&0.736&0.733&0.458\\
			2&0.944&1.000&0.943&0.842&0.736&1.000&0.712&0.458\\
			3&0.947&0.943&1.000&0.834&0.733&0.712&1.000&0.418\\
			4&0.841&0.842&0.834&1.000&0.458&0.458&0.418&1.000\\
			\hline\hline
		\end{tabular}
	\end{table}

	\begin{table}[htbp]
		\caption{The systematic correlation matrices for $D^+\to\eta \mu^+\nu_\mu$.
			\label{tab:etamunua_sysmatrix}}
		\centering
		\begin{tabular}{c|cccc|cccc}
			\hline
			\hline
			& \multicolumn{4}{c|}{$D^+\to\eta_{(2\gamma)}\mu^+\nu_\mu$ }& \multicolumn{4}{c}{$D^+\to\eta_{(3\pi)}\mu^+\nu_\mu$ } \\
			$\epsilon_{ij}$&1&2&3&4&1&2&3&4\\ \hline
			1&1.000&0.912&0.908&0.760&1.000&0.648&0.650&0.388\\
			2&0.912&1.000&0.918&0.772&0.648&1.000&0.646&0.397\\
			3&0.908&0.918&1.000&0.761&0.650&0.646&1.000&0.366\\
			4&0.760&0.772&0.761&1.000&0.388&0.397&0.366&1.000\\
			\hline\hline
		\end{tabular}
	\end{table}
\subsection{Hadronic Form Factor from Separate Fits}

The central values of the fit parameters are taken from the results obtained using the combined
statistical and systematic covariance matrix. The statistical uncertainties of the fit parameters are taken from the fit with only statistical covariance matrix. The systematic uncertainties
are obtained by taking the quadrature difference between the uncertainties from the fit with only statistical covariance matrix and those from the fit with the combined covariance matrix.

The left-side subfigures of Fig.~\ref{fig:ff_etalnu} show the results of the separate fits to the partial decay rates of $D^+\to\eta\ell^+\nu_\ell$. The red dots with error bars are the partial decay rates measured in the data analysis and the black solid curves show the best fits. The right-side subfigures of Fig.~\ref{fig:ff_etalnu} show the projections of the fits on the hadronic form factor as a function of $q^2$, where the dots with error bars show the measured values of the form factor, which are obtained with

\begin{equation} f_+^\eta (q^2_i)=\sqrt{\frac{\Delta\Gamma_i}{\Delta q^2_i}\frac{24\pi^3}{G^2_Fp'^3_\eta(i)|V_{cd}|^2}},
\end{equation}
in which
\begin{equation}
p'^3_\eta(i)=\frac{\int_{q^2_{\rm min,{\it i}}}^{q^2_{\rm max,{\it i}}}p^3_\eta|f_+^\eta(q^2)|^2dq^2}{|f_+^\eta(q_i^2)|^2(q^2_{\rm max,{\it i}}-q^2_{\rm min,{\it i}})}
\end{equation}
where $q^2_{\rm{min,{\it i}}}$ and $q^2_{\rm max,{\it i}}$ are the low and high boundaries of the $i$-th $q^2$ bin. In the calculation, $p'^3_\eta(i)$, $f^\eta_+(q^2)$, and $f^\eta_+(q_i^2)$ are calculated using the two-parameter series expansion model.

Table~\ref{tab:fit_par} gives the obtained results of the fit parameters from the individual fits to
the partial   decay rates of $D^+\to\eta\ell^+\nu_\ell$.

\begin{table}[htbp]
	\centering
	\caption{The parameters of hadronic form factor obtained by fitting the partial decay rates of the semileptonic decays $D^+\to\eta\ell^+\nu_\ell$, where the first uncertainties are statistical and the second systematic, $\rho$ is the correlation coefficient between $r_1$ and $f^\eta_+(0)|V_{cd}|$, and the NDF denotes the number of degrees of freedom.
		\label{tab:fit_par}}
	\resizebox{1.0\textwidth}{!}{
		\begin{tabular}{c|l|c|c|c|c|c}
			\hline\hline
			Case &Decay &$\chi^2$/NDF&$r_1$&$f^\eta_+(0)|V_{cd}|$& $\rho$ &$f^\eta_+(0)$\\
			\hline
			\multirow{4}{*}{Separate fit}
			&$\eta_{(2\gamma)}e^+\nu_e$  &2.6/2&$-3.1\pm0.9\pm0.1$&$0.077\pm 0.002 \pm 0.001$&0.777   &$0.341\pm0.011\pm0.005$ \\
			&$\eta_{(3\pi)}e^+\nu_e$ &0.5/2&$-3.3\pm1.7\pm0.3$&$0.078\pm 0.004 \pm 0.001$ & 0.816  &$0.345\pm0.020\pm0.006$ \\
			&$\eta_{(2\gamma)}\mu^+\nu_\mu$  &0.6/2&$-1.5\pm0.9\pm0.1$&$0.081\pm 0.003 \pm 0.001$&0.819  &$0.359\pm0.013\pm0.006$  \\
			&$\eta_{(3\pi)}\mu^+\nu_\mu$  &1.3/2&$-5.0\pm2.5\pm0.3$&$0.071\pm 0.007 \pm 0.001$  &0.896 &$0.316\pm0.030\pm0.005$  \\\hline
			Simultaneous fit&$\eta\ell\nu_\ell$                  &6.1/14&$-2.8\pm0.7\pm0.2$&$0.078\pm0.002\pm0.001$ & 0.821  &$0.345\pm0.008\pm0.003$ \\
			\hline\hline
		\end{tabular}
		}
\end{table}

\subsection{Hadronic Form Factor from Simultaneous Fit}

To account for the correlation effects in the measurements of the hadronic form factor among the four semileptonic $D^+$ decays, we perform a simultaneous fit to
the partial decay rates of the $\etaenu$ and $\etamunu$ modes to extract the product $\ffeta|V_{cd}|$.

In the simultaneous fit, we still use the least $\chi^2$ method to extract the hadronic form factor.
The $\Delta\Gamma_i$ for the four semileptonic decay modes are combined into one vector of length 16
, and the covariance matrix $C_{ij}$ becomes a $16\times16$ matrix for the simultaneous $\Delta\Gamma_i$ fit, which is redefined as $C_{ij}=C^{\rm stat}_{ij}+C^{\rm csyst}_{ij}+C^{\rm usyst}_{ij}$, $(i,j=1,2,3,...,15,16)$, where $C^{\rm stat}_{ij}$ is the statistical covariance matrix defined as
\begin{linenomath*}
	$$C^{\rm stat}_{ij}=
	\begin{pmatrix}
		A_1&0&0&0\\
		0&A_2&0&0\\
		0&0&B_1&0\\
		0&0&0&B_2
	\end{pmatrix},
	$$
\end{linenomath*}
where $A_1$, $A_2$, $B_1$, and $B_2$ represent the statistical covariance matrices obtained for the individual signal decays.
The $C^{\rm csyst}_{ij}$ is the correlated systematic covariance matrix given by
\begin{equation}
	\label{eq:csys_matrix}
	C_{ij}^{\rm{csyst}}=\delta(\Delta\Gamma_i)\delta(\Delta\Gamma_j).
\end{equation}

For the uncorrelated systematic uncertainties, the systematic covariance matrix $C^{\rm usyst}_{ij}$ is defined as
\begin{linenomath*}
	$$C^{\rm usyst}_{ij}=
	\begin{pmatrix}
		a_1&0&0&0\\
		0&b_1&0&0\\
		0&0&c_1&0\\
		0&0&0&d_1
	\end{pmatrix}
	,
	$$
\end{linenomath*}
where $a_1$, $b_1$, $c_1$, and $d_1$ represent the uncorrelated systematic covariance matrices obtained from individual semileptonic decays.

\begin{figure}[htbp]
	\begin{center}
		\includegraphics[width=1.0\textwidth]{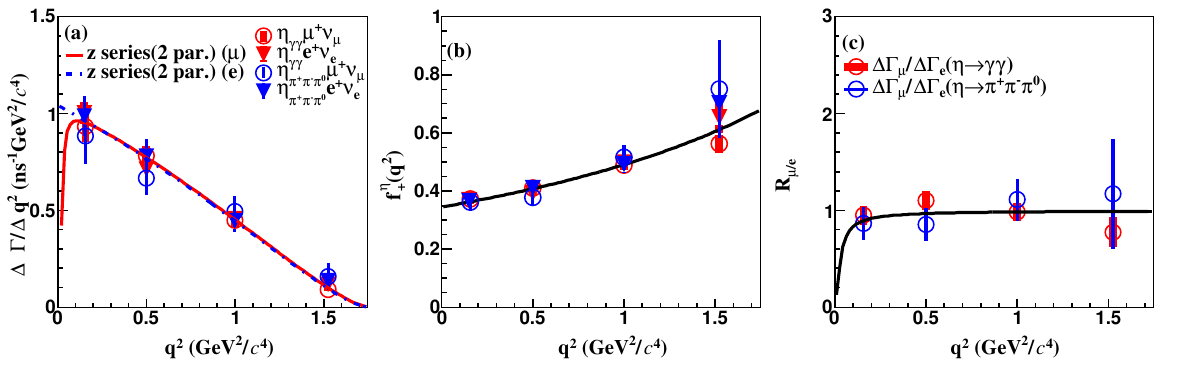}
		\caption{(a) Simultaneous fit to partial decay rates of $\etalnu$. (b) Projection to the form factor as function of $q^2$ of $\etalnu$, and the black curve is the fit result. (c) The measured $R_{\mu/e}$ in each $q^2$ interval, and the black curve is the SM prediction.
			\label{ff_etalnu}}
	\end{center}
\end{figure}

A simultaneous fit is performed on the partial decay rates of the four semileptonic $D^+$ decays, using the modified $\Delta\Gamma_i$ and $C_{ij}$,
In the fit, the four semileptonic $D^+$ decays share the same hadronic form factor parameters.
Figure~\ref{ff_etalnu}(a) exhibits the fit result, Fig.~\ref{ff_etalnu}(b) shows the extracted hadronic form factor, and Fig.~\ref{ff_etalnu}(c) displays the measured ratios of partial decay rates of $D^+\to\eta\mu^+\nu_{\mu}$ and $D^+\to\eta e^+\nu_e$, $R_{\mu/e}=\Delta\Gamma_\mu/\Delta\Gamma_e$, in each $q^2$ interval.
The fitted parameters obtained from this simultaneous fit are summarized in the last row of Table~\ref{tab:fit_par}.

\section{SUMMARY}

In summary, by analyzing 20.3 fb$^{-1}$ of $e^+e^-$ collision data collected at $\sqrt{s}=3.773$~GeV with the BESIII detector, we report precision measurements of the semileptonic decays $\etaenu$ and $\etamunu$. The absolute BFs of these decays are determined to be
\begin{equation}
	{\mathcal B}(\etaenu) = (9.75\pm0.29\pm0.28)\times10^{-4},
	\nonumber
\end{equation}
\begin{equation}
	{\mathcal B}(\etamunu) = (9.08\pm0.35\pm0.23)\times10^{-4},
	\nonumber
\end{equation}
both of which improve over the previous best measurements by more than two fold in precision.
Combining the BFs of semielectronic and semimuonic decays, we obtain the ratio
\begin{equation}
	\frac{\mathcal B(\etamunu)}{\mathcal B(\etaenu)}=0.93\pm0.05\pm0.02,
	\nonumber
\end{equation}
where the correlated uncertainties cancel. This result is well consistent with theoretical calculation~\cite{Rbf1,Rbf2}, thereby implying no violation of lepton universality under the current statistics.

From the simultaneous fit to the partial   decay rates of the semileptonic decays $\etaenu$ and $\etamunu$, we determine the product of the hadronic form factor $\ffeta$ and the CKM matrix element $|V_{cd}|$ to be
\begin{equation}
	\ffeta|V_{cd}|=0.078\pm0.002\pm0.001.
	\nonumber
\end{equation}
Taking the $|V_{cd}|=0.22486\pm0.00067$ given by the PDG~\cite{pdg2022} as input, we obtain the hadronic form factor to be
\begin{equation}
	\ffeta=0.345\pm0.008\pm0.003.
	\nonumber
\end{equation}

\begin{figure}[htp]\centering
	\setlength{\abovecaptionskip}{-2pt}
	\setlength{\belowcaptionskip}{-3pt}
	\includegraphics[width=0.6\textwidth]{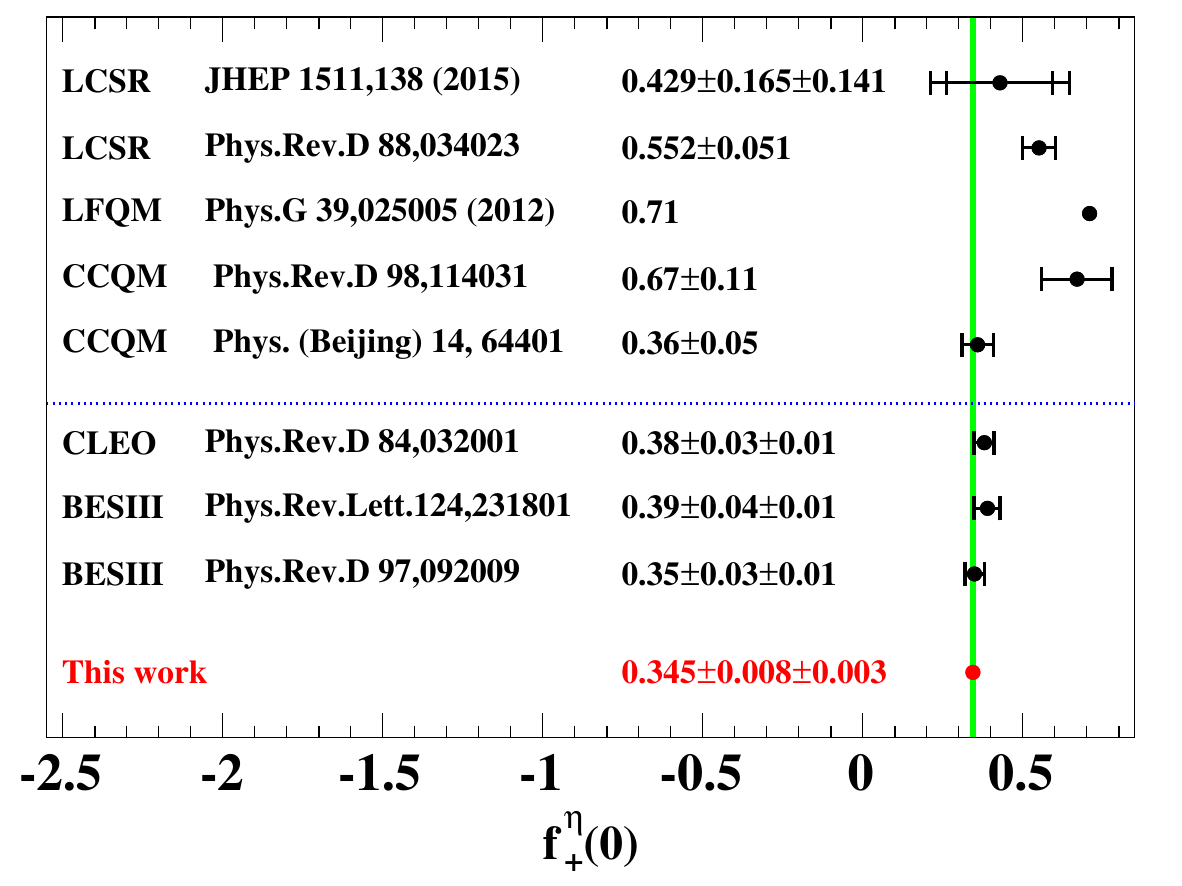}
	\caption{Comparison of $f^\eta_+(0)$ measured in this analysis with theoretical calculations.
		The first and second uncertainties are statistical and systematic, respectively.  The green band corresponds to the $\pm1\sigma$ limit of our measurement.
		\label{compare_ff_etalnu}
	}
\end{figure}

Figure~\ref{compare_ff_etalnu} shows the comparison of the $\ffeta$ obtained in this analysis with the previous measurements and theoretical predictions.
Our meausured $\ffeta$ is consistent with the previous measurements, and the precision is improved by a factor of 3.4 over the previous best one.
This is important for testing different theoretical models.

\section{ACKNOWLEDGMENTS}
The BESIII Collaboration thanks the staff of BEPCII and the IHEP computing center for their strong support. This work is supported in part by National Key R\&D Program of China under Contracts Nos. 2023YFA1606000, 2023YFA1606704; National Natural Science Foundation of China (NSFC) under Contracts Nos. 11635010, 11735014, 11935015, 11935016, 11935018, 12025502, 12035009, 12035013, 12061131003, 12192260, 12192261, 12192262, 12192263, 12192264, 12192265, 12221005, 12225509, 12235017, 12361141819; the Chinese Academy of Sciences (CAS) Large-Scale Scientific Facility Program; the CAS Center for Excellence in Particle Physics (CCEPP); Joint Large-Scale Scientific Facility Funds of the NSFC and CAS under Contract No. U1832207; CAS under Contract No. YSBR-101; 100 Talents Program of CAS; The Institute of Nuclear and Particle Physics (INPAC) and Shanghai Key Laboratory for Particle Physics and Cosmology; Agencia Nacional de Investigación y Desarrollo de Chile (ANID), Chile under Contract No. ANID PIA/APOYO AFB230003; German Research Foundation DFG under Contract No. FOR5327; Istituto Nazionale di Fisica Nucleare, Italy; Knut and Alice Wallenberg Foundation under Contracts Nos. 2021.0174, 2021.0299; Ministry of Development of Turkey under Contract No. DPT2006K-120470; National Research Foundation of Korea under Contract No. NRF-2022R1A2C1092335; National Science and Technology fund of Mongolia; National Science Research and Innovation Fund (NSRF) via the Program Management Unit for Human Resources \& Institutional Development, Research and Innovation of Thailand under Contract No. B50G670107; Polish National Science Centre under Contract No. 2019/35/O/ST2/02907; Swedish Research Council under Contract No. 2019.04595; The Swedish Foundation for International Cooperation in Research and Higher Education under Contract No. CH2018-7756; U. S. Department of Energy under Contract No. DE-FG02-05ER41374

\bibliographystyle{JHEP}

\end{document}